\documentclass[useAMS,usenatbib]{mn2e}
\usepackage{url}
\usepackage{hyperref}
\usepackage[dvips]{graphicx}
\usepackage{float}
\usepackage{xcolor}
\def\mnras{{MNRAS}}
\def\apj{{ApJ}}
\def\aj{{AJ}}
\def\aap{{A\&A}}
\def\apjl{{ApJL}}

\def\pasp{{PASP}}

\def\procspie{{Proc. SPIE}}

\def\ecs{ergs cm$^{-2}$ s$^{-1}$}
\def\ecsa{ergs cm$^{-2}$ s$^{-1}$ \AA$^{-1}$}

\def\msun{$M_{\odot}$}

\def\me{{$\dot{m}_{E}$}}

\def\ltsim{\mathrel{\hbox{\rlap{\hbox{\lower4pt\hbox{$\sim$}}}\hbox{$<$}}}}
\def\gtsim{\mathrel{\hbox{\rlap{\hbox{\lower4pt\hbox{$\sim$}}}\hbox{$>$}}}}
\def\rs{{$r_{s}$}}

\def\rg{{$R_{g}$}}
\def\rout{{$R_{out}$}}
\def\rsub{{$R_{sub}$}}
\def\us{{$u_{s}$}}
\def\gs{{$g_{s}$}}
\def\rs{{$r_{s}$}}
\def\is{{$i_{s}$}}
\def\zs{{$z_{s}$}}

\begin{document}

\title[First detection of AGN accretion disc outer edge]{First detection of the outer edge of an AGN accretion disc: Very fast multiband optical variability of NGC~4395 with GTC/HiPERCAM and LT/IO:O.
}
\author[M$\rm^{c}$Hardy, I.M.]
{ I M M$\rm^{c}$Hardy$^{1}$, M Beard$^{1}$, E Breedt$^{2}$, J H Knapen$^{3,4}$, 
F M Vincentelli$^{1,3,4}$, \and  M Veresvarska$^{1,5}$,  
V S Dhillon$^{3,6}$, T R Marsh$^{7}$, S P Littlefair$^{6}$,
K Horne$^{8}$,  R Glew$^{1}$, \and M R Goad$^{9}$, E Kammoun$^{10,11}$
and D Emmanoulopoulos$^{1}$\\
$^{1}$ Department of Physics and Astronomy, The University, Southampton
SO17 1BJ\\
$^{2}$ Institute of Astronomy, Madingley Road, Cambridge, CB3 0HE\\
$^{3}$ Instituto de Astrof\'{i}sica de Canarias, V\'{i}a L\'{a}ctea S/N, E-38205 La Laguna, Spain \\
$^{4}$ Departamento de Astrof\'{i}sica, Universidad de La Laguna, E-38206 La Laguna, Spain\\
$^{5}$ Department of Physics, University of Durham, South Road, Durham DH1 3LE\\
$^{6}$ Dept of Physics and Astronomy, University of Sheffield, Sheffield S3 7RH\\
$^{7}$ Department of Physics, University of Warwick, Coventry CV4 7AL\\
$^{8}$ School of Physics and Astronomy, University of St Andrews, North Haugh, St Andrews KY16 9SS \\
$^{9}$ Department of Physics and Astronomy, University Rd, Leicester LE1 7RH \\
$^{10}$ IRAP, Universit\'e de Toulouse, CNRS, UPS, CNES 9, Avenue du Colonel Roche, BP 44346, F-31028, Toulouse Cedex 4, France\\
$^{11}$ INAF - Osservatorio Astrofisico di Arcetri, Largo Enrico Fermi 5, I-50125 Firenze, Italy
}

\date{Accepted 2022 December 5. Received 2022 December 4; in original form 2022 May 4}
\maketitle

\begin{abstract}
We present fast ($\sim$200s sampling) $\it ugriz$ photometry of the low mass AGN NGC\,4395 with the Liverpool Telescope, followed by very fast (3s sampling) \us,\gs,\rs,\is\, and \zs\ simultaneous monitoring with HiPERCAM on the 10.4m GTC. These observations provide the fastest ever AGN multiband photometry and very precise lag measurements.
Unlike in all other AGN, \gs\, lags \us\, by a large amount, consistent with disc reprocessing but not with reprocessing in the Broad Line Region (BLR). There is very little increase in lag with wavelength at long wavelengths, indicating an outer edge (\rout) to the reprocessor. We have compared truncated disc reprocessing models to the combined HiPERCAM and previous X-ray/UV lags. For the normally accepted mass of $3.6 \times 10^{5}$\msun\, we obtain reasonable agreement with zero spin, \rout$\sim 1700$\rg\, and the {\sc DONE} physically-motivated temperature-dependent disc colour correction factor (f$\rm_{col}$). 
A smaller mass of $4 \times 10^{4}$\msun\, can only be accomodated if
f$\rm_{col}=2.4$, which is probably unrealistically high.
Disc self gravity is probably unimportant in this low mass AGN but an obscuring wind may provide an edge. For the small mass the dust sublimation radius is similar to \rout\, so the wind could be dusty.
However for the more likely large mass the sublimation radius is further out so the optically-thick base of a line-driven gaseous wind is more likely. The inner edge of the BLR is close to \rout\, in both cases. These observations provide the first good evidence for a truncated AGN disc and caution that truncation should be included in reverberation lag modelling.
\end{abstract}


\section{Introduction}
\label{sec:intro}

The origin of UV/optical variations in AGN, their relationship to X-ray variations and what those variations together can tell us about the inner structure of AGN, has been a matter of major observational activity for over two decades. Initially, combined X-ray monitoring from RXTE and optical monitoring from the ground \cite[e.g.][]{uttley03_5548, suganuma06, arevalo08_2251, arevalo09, breedt09, breedt10, breedt10_thesis, lira11} typically showed optical variations lagging the X-rays by about a day ($\pm 0.5$d). These observations are consistent with reprocessing of X-rays from around the central black hole by a surrounding accretion disc, but usually without sufficient multiband detail to map out the temperature structure of the disc \citep{blandford_mckee82}. Some multiband optical monitoring \cite[e.g.][]{sergeev05, cackett07} showed wavelength-dependent lags consistent with the expectations of reprocessing from an accretion disc but without simultaneous X-ray monitoring. More recently, X-ray and UV/optical multiwaveband monitoring, based mainly around Swift observations \cite[e.g.][]{cameron12, cameron14_thesis, shappee14, mch14, edelson15, troyer16, fausnaugh16, mch18, cackett18, edelson19, cackett20, hernandez20, vincentelli21, vincentelli22}, but also with XMM-Newton \cite[e.g.][]{mch16},
have confirmed that the UV/optical variations are mostly well explained by reprocessing of high energy radiation from an accretion disc, but have also noted a reprocessed contribution from the gas in the broad line region (BLR). 
The lags expected from the BLR were predicted by \cite{korista01, korista19} and 
are particularly large in the $u$ band, where combined Balmer emission lines form the Balmer continuum, and also in the $i$ band from the Paschen continuum \cite[see particularly][]{cackett18}. 
The combined signatures of reprocessed disc and BLR emission are clearly seen in NGC~4593
\citep{mch18}. Here the response functions \citep{horne04} required to explain the UV and optical lightcurves by reprocessing of X-rays, show both a short-timescale (disc) peak and long-timescale (BLR) tail.  In Mrk110, \cite{vincentelli21} also show a combination of short-timescale (disc) and long-timescale (BLR) lags.

Given the many observations and many suggested explanations, there is increasing interest in modelling the lag spectra. Initially most modelling was based on reprocessing of X-rays by a disc with the temperature profile described either by \cite{shakura73} or (with GR corrections) by \cite{novikov73}. 
Although the broad reprocessing picture is now established, this modelling highlighted a variety of problems
including:\\ 
\noindent
(1) The observed optical lightcurves are much smoother than predicted \cite[e.g][]{berkley00,arevalo08_2251} unless the X-ray emitting corona is much larger ($\sim100$ gravitational radii, $R_{g}$) than estimated from X-ray reverberation \citep[$\sim4 R_{g}$, e.g.][]{emmanoulopoulos14,cackett14} or microlensing observations \citep[$\ltsim 10 R_{g}$][]{dai10,chartas12,mosquera13}\\ \noindent (2) The disc sizes estimated from almost all Swift monitoring are larger than predicted,  with analytic formulae \citep[e.g.][]{fausnaugh16} giving larger discrepancies ($>3 \times$ larger) than numerical disc modelling \citep[$\ltsim2 \times$,][]{mch18}. However numerical modelling which better includes relativistic effects \citep{kammoun21_fits} gives no discrepancy, albeit with an X-ray source height ($\sim 20-70~R_{g}$) much larger than that of most previous modelling.\\ 
\noindent(3) The lag between the X-ray and far UV band is much larger than expected, compared with the lags between the various UV and optical bands \cite[e.g. summary in][]{mch18}. Possible solutions include scattering through an inflated inner disc \citep{gardner17} or including a more distant reprocessor such as the BLR \cite{mch18}. Indeed \citet{netzer22} claims that the BLR dominates the lags.
Alternatively, if all lags are measured relative to a reference UV band, then the X-ray to UV lag may be increased if the X-ray auto-correlation function is broad \citep{kammoun21_model}.

Reprocessing, of course, cannot explain all aspects of UV/optical variability. There are long term (months-years) trends in the UV/optical which are not mirrored in the X-rays and which may result from inwardly propagating accretion rate fluctuations in the disc \cite[e.g.][]{arevalo08_2251, breedt09, breedt10}. If these trends are filtered out then the shorter-timescale reprocessing signature can be revealed \citep{mch14,mch18,pahari20}. However in this paper we are concentrating on the short-timescale variations.

Most AGN which have been monitored so far have been of similar black hole masses, mostly between $\sim10^{7}$ and $\sim10^{8}$ \msun. Although accretion rates are notoriously hard to measure accurately, the majority of AGN monitored so far have had accretion rates of few per cent of Eddington, although a small number have higher rates, eg  Mrk110,  \me  $\sim40$ per cent \citep{vincentelli21},  NGC7469, \me  $\sim50$ per cent \citep{pahari20} and Mrk142 \me $\sim100$ per cent \citep{cackett20}. There have been very few observations of very low mass or very low accretion rate AGN, largely because most of them are too faint. However to test models, it is important to extend the range and hence objects at the extreme edges of the distribution, where disc and BLR properties are likely to be most different, are particularly useful. For example, at very low accretion rates, the disc scale height is likely to be lower \citep{treves88} thus affording a less obscured view to the outer disc.

Here we present lag measurements of NGC~4395. Although its exact mass is currently a matter of some debate, with values between $3.6 \times 10^{5}$\msun\, \citep{peterson05} and $1 \times 10^{4}$\msun\, \citep{woo19} having been claimed, there is complete agreement that its mass is definitely very low. For the mass noted by 
\cite{peterson05}, the accretion rate is also very low, 0.12 per cent Eddington so its luminosity is low. However it is close enough to us \citep[3.85 Mpc,][]{tully09} that it is bright enough for accurate lag measurements to be made.

Early Swift observations \citep{cameron12} showed that the B-band lagged the X-rays in NGC~4395 by less than 45~min, with a tentative estimate of the UVW2 band lagging the X-rays by 400s. XMM-Newton observations, using the optical monitor in fast readout mode, showed that the UVW1 band lagged the X-rays by $473^{+47}_{-98}$s and parallel ground based observations gave a lag of the X-rays by the $g$ band of $788^{+44}_{-54}$s. There have, however, been no simultaneous multiband lag measurements and so serious comparison with reprocessing models has not yet been possible. 

In Section 2 we discuss {\it ugriz} monitoring of NGC~4395 with the robotic 2m Liverpool Telescope (LT) which showed variability on few hour timescales, consistent with the few hundred second lags noted above, but were not able to measure the lags accurately. However based on those observations we obtained very high sensitivity simultaneous  {\it ugriz} photometry over one night with HiPERCAM on the 10.4m Gran Telescopio Canarias (GTC) with time resolution down to 3s (Section 3), thus allowing very accurate measurement of interband lags. In Section 4 we compare the resultant lag measurements with reprocessing models. Conclusions are presented in Section 5.

\section{Liverpool Telescope Observations}

\subsection{The Observations}

NGC~4395 was observed with the IO:O instrument on the robotic Liverpool Telescope \cite[LT,][]{steele04} for approximately 6 hours on each of 5 separate nights in 2017 March.
The IO:O\footnote{\url{https://telescope.livjm.ac.uk/TelInst/Inst/IOO}}
is a CCD imaging camera with a $10 \times 10$ arcmin field of view and an unbinned pixel size of 0.15 arcsec, here used in $2 \times 2$ binning mode. 

\begin{table*}
\begin{tabular}{llllll}
Night & Date (UTC) & Start Time & End Time & Filters and Exposure Times & Lightcurve Figure \\
&&&&&\\
1 &2017-03-23 & 00:49:05 & 03:23:10 & {\it g,r,i,z}  all 10s & Fig.~\ref{fig:griz_sec1}\\
2 &2017-03-24 & 01:03:46 & 07:01:45 & {\it u}-30s, {\it B}-20s, {\it g}-20s & Fig.~\ref{fig:ubg_sec2}\\
3 &2017-03-26 & 00:40:01 & 06:39:45 & {\it g,r,i,z}  all 20s & Fig.~\ref{fig:griz_sec3}\\
4 &2017-03-28 & 00:34:54 & 06:31:33 & {\it u}-30s, {\it B}-20s, {\it g}-20s &  Fig.~\ref{fig:ubg_sec4}\\
5 &2017-03-29 & 00:33:37 & 06:32:06 & {\it g,r,i,z}  all 20s & Fig.~\ref{fig:griz_sec5}
\end{tabular}
\caption{Log of LT observations of NGC~4395. Start Time is the start time of the first observation in the series and End Time is the start time of the last observation.}
\label{tab:ltlog}
\end{table*}

The observations were made by cycling through successive exposures in groups of filters. Given the previously measured approximate lag of 330s between the UVW1 (290 nm) and $g$ band \citep{mch16}, the observations were arranged in two groups so that cycle times between the same filter did not exceed 200s, thus allowing reasonable sampling of likely lag times. 
To allow measurement of lags in all bands relative to the $g$-band, one group consisted of the {\it g, r, i} and {\it z} bands and the other consisted of the {\it g, B} and {\it u} bands. 
The {\it g, r, i, z} group was observed on nights 1, 3 and 5. 
The {\it g, B, u} group was observed on nights 2 and 4. Due to bad weather, these nights were not consecutive.
The log of the observations is given in Table~\ref{tab:ltlog}. 

To remove any non-astrophysical variations, such as those due to changes in atmospheric transparency which would affect all objects on a CCD image, lightcurves were produced using the standard method of differential photometry relative to one or more comparison stars.

\begin{figure}
\includegraphics[width=80mm,height=100mm,angle=0]{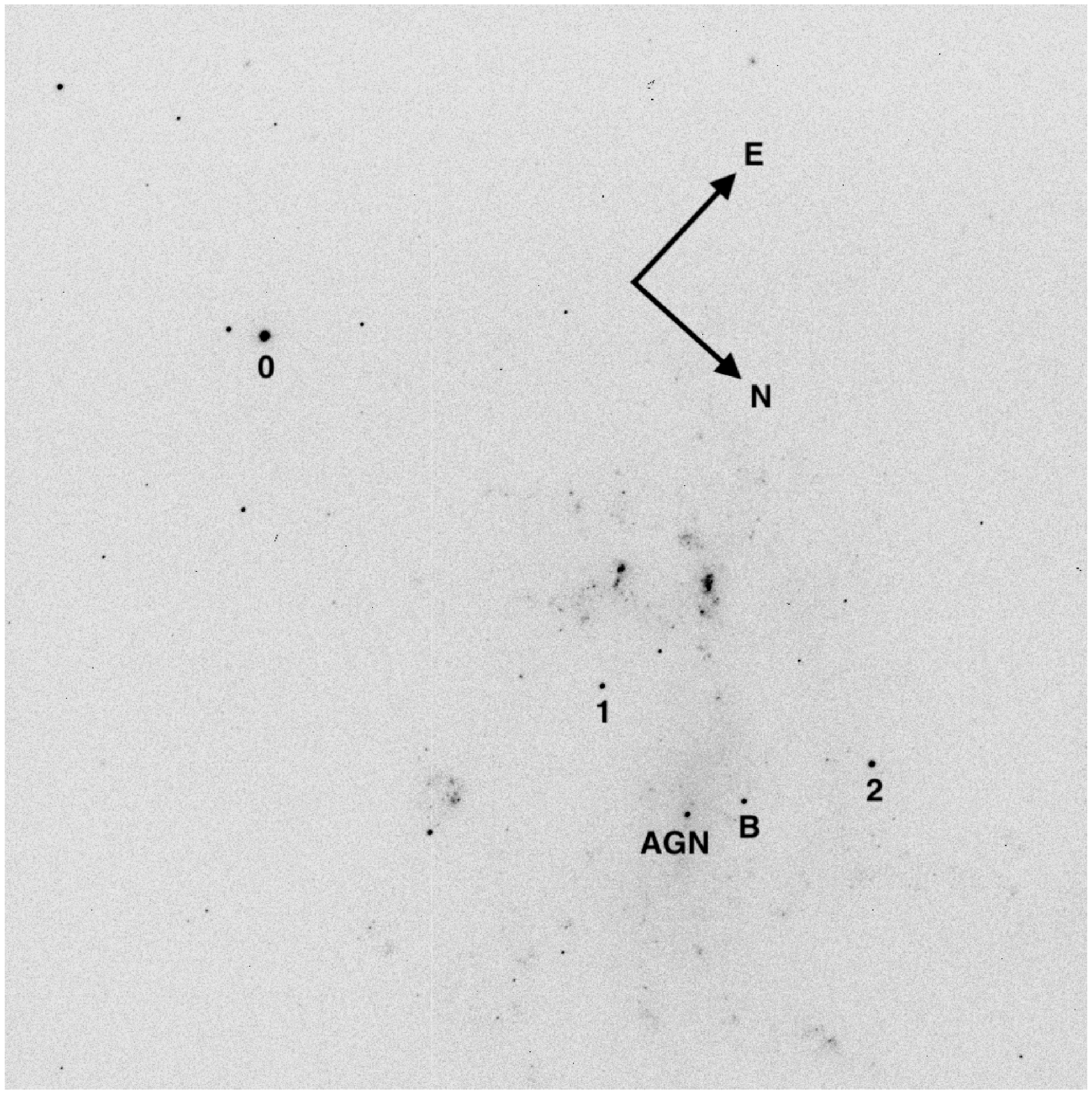}
\vspace*{-10mm}
\caption{$g$-band image of NGC~4395, 40s exposure, with the IO:O CCD imager on the Liverpool Telescope. The comparison stars, 0 and 2, used in the production of the LT lightcurves are labelled, as are the comparison stars, 1 and B, used in the production of the HiPERCAM lightcurves. The AGN nucleus is also labelled. The field of view is $10 \times 10$ arcmin.}
\label{fig:lt_finder}
\end{figure}

An LT IO:O $g$-band image of NGC~4395 is shown in Fig.~\ref{fig:lt_finder}. In this figure a number of comparison stars are labelled, together with the AGN nucleus. Fig.~\ref{fig:lt_finder} is centred midway between comparison stars 0 
(RA 12 25 54.63  Dec +33 26 51) and 2 (RA 12 25 55.87  Dec  +33 33 58). However the observations reported in this paper were centred on the AGN. Star 2 was always on the field of view, well away from the edge of the CCD, but star 0 was outside of the field of view. Star 2 was therefore used as the comparison star. Also labelled in 
Fig.~\ref{fig:lt_finder} are stars 1 and B which are within the smaller HiPERCAM field of view. 

For each CCD image, an aperture of radius 4.25 arcseconds, which is 14 LT (binned) pixels, was placed around the AGN and a similar aperture was placed around star 2. A background annulus of inner radius $3 \times$ the source radius and outer radius $4 \times$ was also placed around each object. 
Lightcurves were derived as the ratio of the background subtracted counts in the AGN compared to those of the comparison star.

\begin{figure}
\hspace*{-10mm} 
\includegraphics[width=70mm,height=90mm,angle=270]{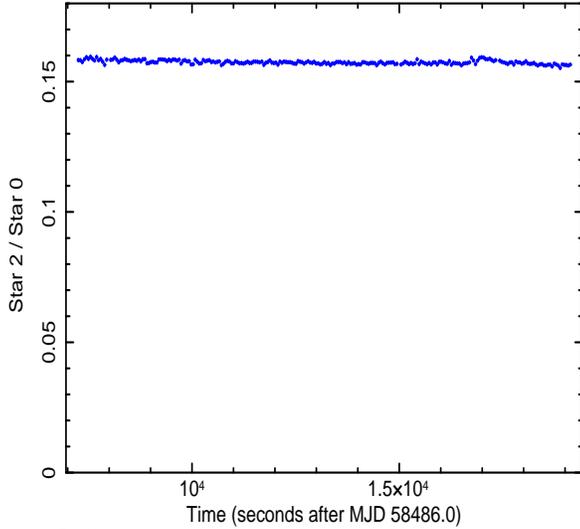}
\vspace*{-2mm}
\caption{Ratio of the background subtracted count rate of star 2 compared to that of star 0 in the g-band. From Beard et al (in prep).}
\label{fig:comparison}
\end{figure}

In later observations which will be reported elsewhere (Beard et al, in prep), both stars 0 and 2 were on the IO:O field of view. In Fig.~\ref{fig:comparison} we show one of the IO:O lightcurves of star 2 compared to star 0. The middle half of this lightcurve is constant (reduced $\chi^{2}=0.94$) but including the ends there is a very slight fractional decrease of 0.006. Stars 0 and 2 have different colours. From SDSS magnitudes g-r = 1.33 for star 2 and 0.98 for star 0. We have not corrected the lightcurves presented here for colour related changes in atmospheric transmission during these observations but it is quite likely that this very small change might be explained in that way. We therefore conclude that both stars are non-variable over the timescales of the observations reported here and can be used safely as comparison stars. Note that the zero point in Fig.~\ref{fig:comparison} is zero in order to demonstrate the relative constancy of the ratio. 

We performed differential photometry using two different pipelines. The first pipeline was an in-house Southampton University pipeline and the second is the HiPERCAM pipeline.\footnote{\url{http://deneb.astro.warwick.ac.uk/phsaap/hipercam/docs/html/}}
Both pipelines use very similar methods, centroiding on the objects of interest and then performing aperture photometry using a background annulus surrounding the object. The results were very similar. Here we present the results using the HiPERCAM pipeline, in large part because this pipeline is well established and well documented and the code is already publicly available thus allowing future researchers to check our results more easily.\\

\vspace*{-2mm}

\vspace*{-2mm}

\begin{figure}
\includegraphics[width=80mm,height=90mm,angle=270]{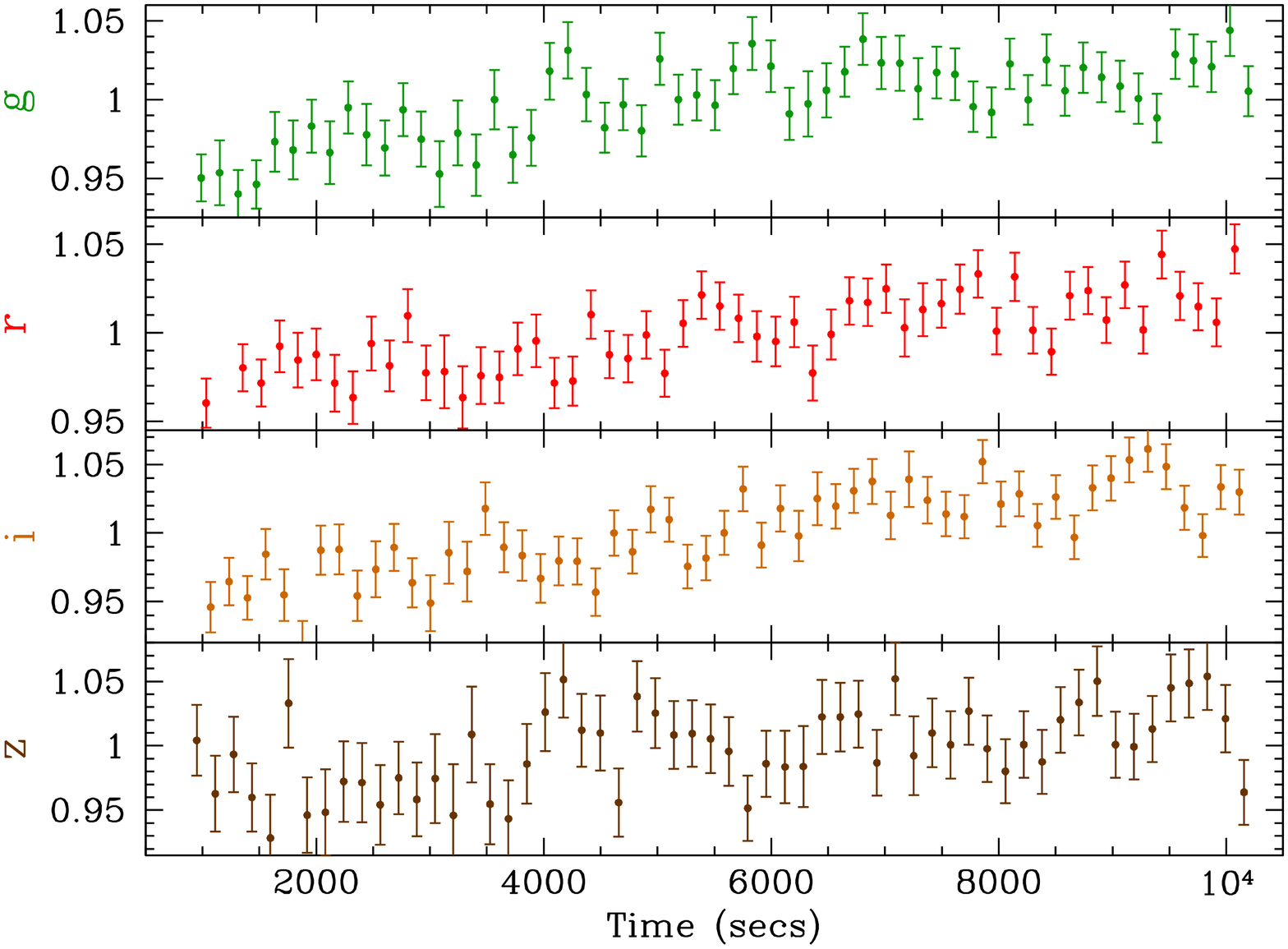}
\vspace*{-2mm}
\caption{Night 1: Lightcurves of NGC~4395 in, from the top, the {\it g, r, i} and {\it z} bands, from the Liverpool Robotic Telescope. Fluxes are normalised to the mean. UT times are given in Table~\ref{tab:ltlog}. }
\label{fig:griz_sec1}
\end{figure}
\vspace*{-2mm}

\begin{figure}
\includegraphics[width=80mm,height=90mm,angle=270]{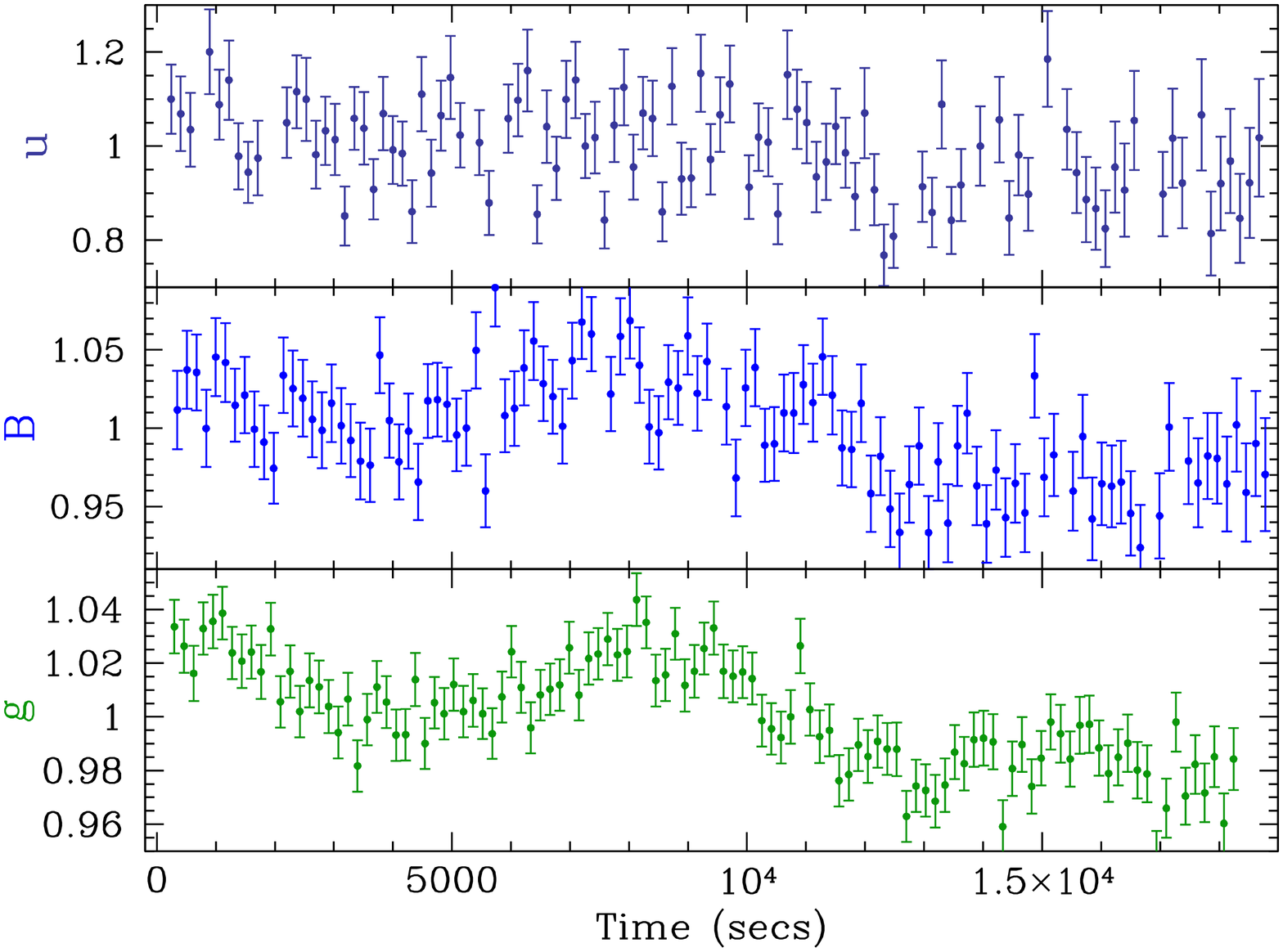}
\vspace*{-2mm}
\caption{Night 2: Lightcurves of NGC~4395 in, from the top, the {\it u, B} and {\it g} bands from the Liverpool Robotic Telescope. Fluxes are normalised to the mean. UT times are given in Table~\ref{tab:ltlog}.
}
\label{fig:ubg_sec2}
\end{figure}
\vspace*{-2mm}

\begin{figure}
\includegraphics[width=80mm,height=90mm,angle=270]{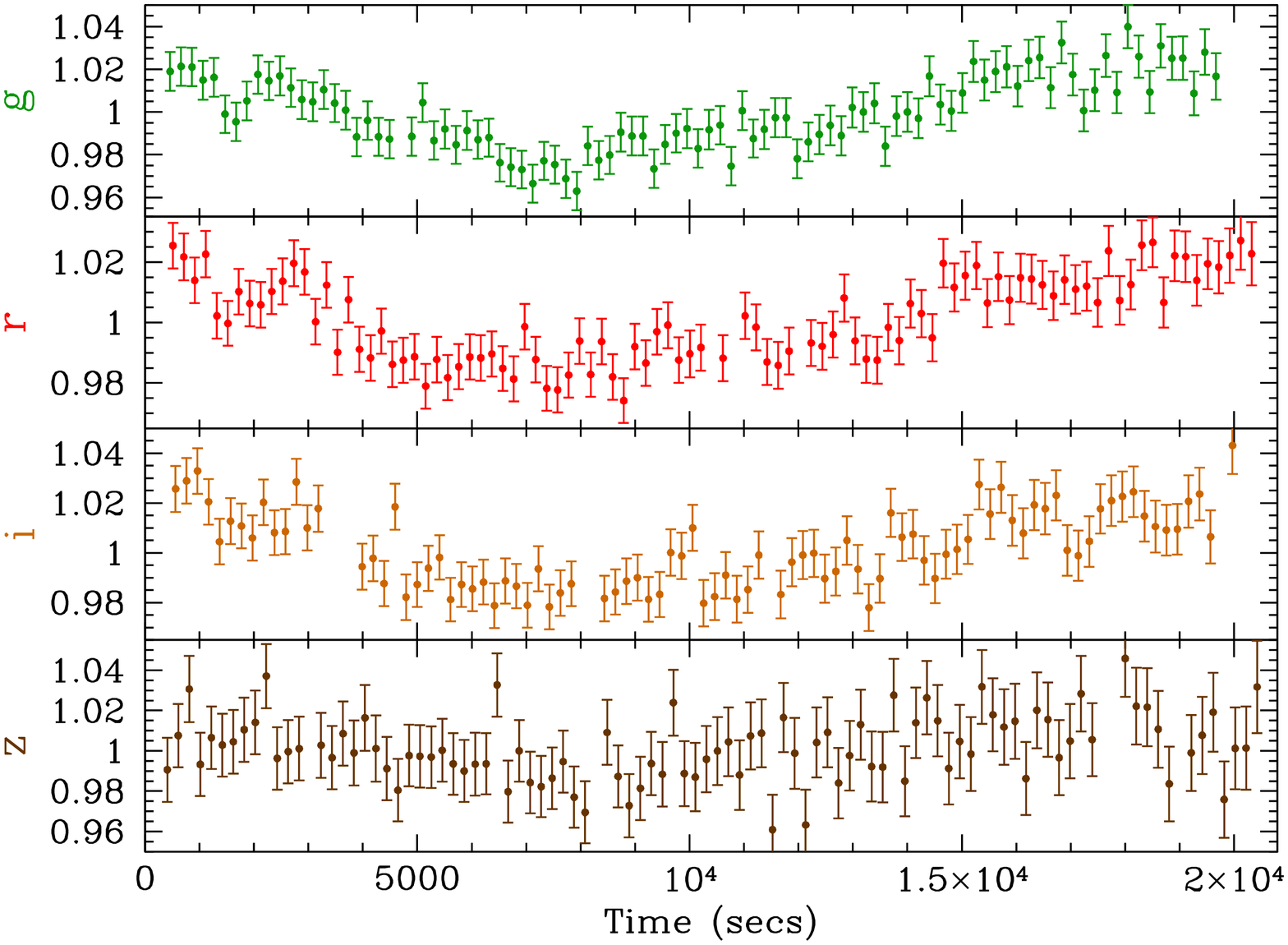}
\vspace*{-2mm}
\caption{Night 3:
As for Fig.~\ref{fig:griz_sec1}.
}
\label{fig:griz_sec3}
\end{figure}
\vspace*{-2mm}

\begin{figure}
\includegraphics[width=80mm,height=90mm,angle=270]{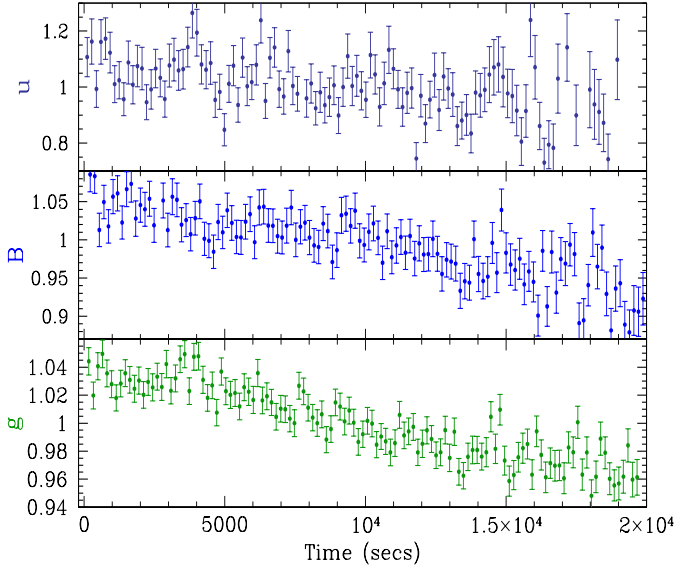}
\vspace*{-2mm}
\caption{Night 4:
As for Fig.~\ref{fig:ubg_sec2}.
}
\label{fig:ubg_sec4}
\end{figure}
\vspace*{-2mm}

\begin{figure}
\includegraphics[width=80mm,height=90mm,angle=270]{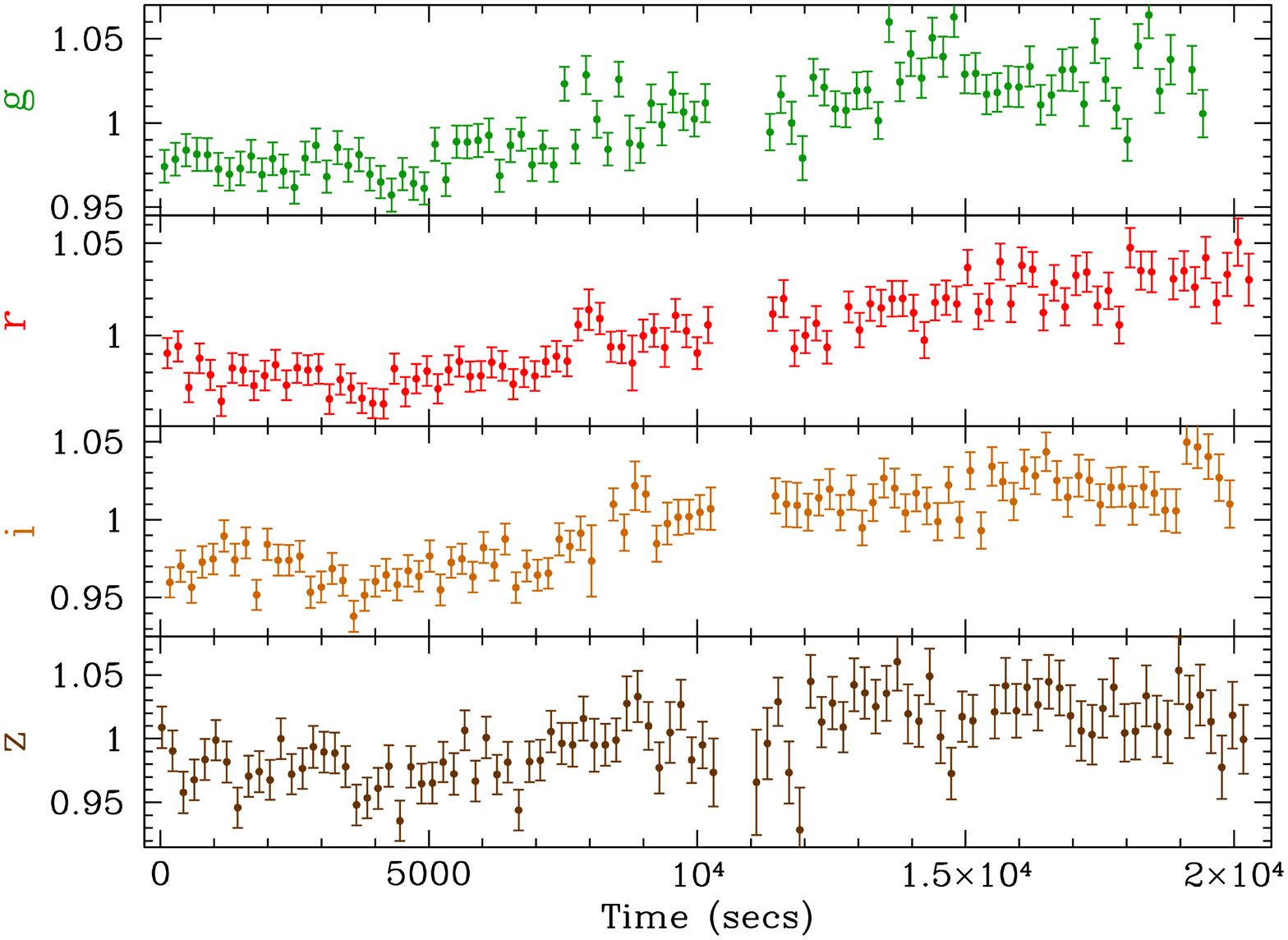}
\vspace*{-2mm}
\caption{Night 5:
As for Fig.~\ref{fig:griz_sec1}.
}
\label{fig:griz_sec5}
\end{figure}

\vspace*{5mm}

\subsection{Interband Correlations and Lags}

The lightcurves for nights 1, 2, 3, 4 and 5 are shown in Figs.~\ref{fig:griz_sec1}, ~\ref{fig:ubg_sec2}, ~\ref{fig:griz_sec3}, ~\ref{fig:ubg_sec4} and ~\ref{fig:griz_sec5} respectively. Here, unlike in  Fig.~\ref{fig:comparison}, the ordinate does not start at zero. Note that the fluxes typically vary by $\sim10$ per cent, which is much larger than the 0.6 per cent trend in the comparison stars.
In all cases the individual lightcurves on each night had similar shapes and are well correlated. However on nights 1 and 5, the {\it griz} lightcurves were dominated by slow trends, without enough prominent features to measure a lag between the bands. Similarly, the {\it ubg} lightcurves on night 4 were also dominated by slow trends. However the {\it griz} lightcurves on night 3 and the {\it ubg} lightcurves on night 2 did show more features. 

\begin{figure}
\hspace*{-5mm}
\includegraphics[width=60mm,height=85mm,angle=270]{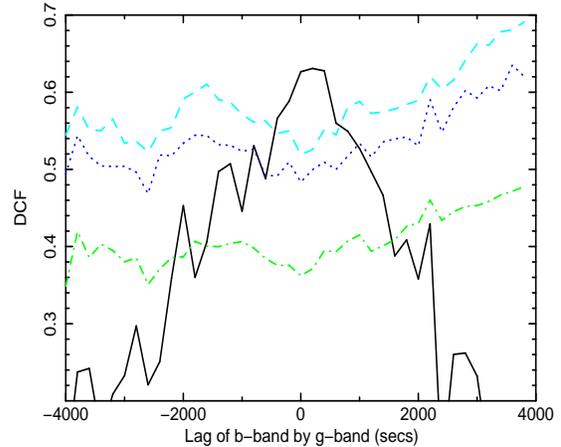}
\vspace*{-2mm}
\caption{DCF showing the lag of the B-band by the $g$ band in seconds for LT night 2 (Fig.~\protect\ref{fig:ubg_sec2}).  From bottom to top, the 68, 90 and 95 per cent confidence contours are shown.
}
\label{fig:bg_dcf_sims}
\end{figure}

\begin{figure}
\hspace*{-5mm} 
\includegraphics[width=60mm,height=85mm,angle=270]{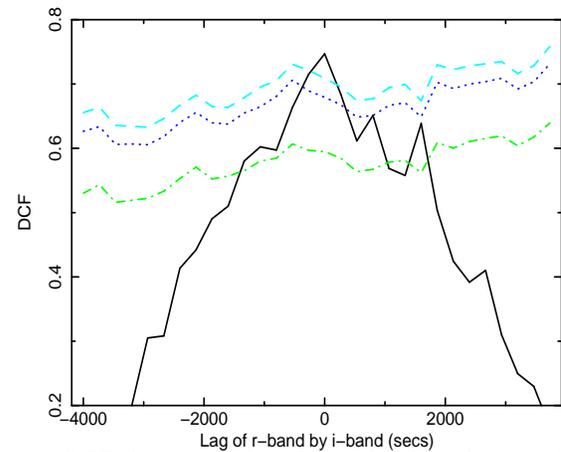}
\vspace*{-2mm}
\caption{DCF showing the lag of the $r$ band by the $i$ band in seconds for LT night 3 (Fig.~\protect\ref{fig:griz_sec3}). From bottom to top, the 68, 90 and 95 per cent confidence contours are shown.
}
\label{fig:ri_dcf_sims}
\end{figure}

In Figs.~\ref{fig:bg_dcf_sims} and ~\ref{fig:ri_dcf_sims} we show, as representative, the simple discrete correlation functions \citep[DCFs:][]{edelson88} between the B and $g$ bands on night 2 and between the $r$ and $i$ bands on night 3. We also show the 68, 90 and 95 per cent confidence contours. The observed correlations easily exceed 95 per cent confidence and so are unlikely to have occured by chance. The confidence contours are generated as described in \cite{breedt09} and \cite{mch18}. This process involves simulating 1000 realisations of the shorter wavelength lightcurve based on its power spectrum. Specifically here we note that as the power spectrum derived from the observed lightcurve is not of sufficient quality to measure accurately its low and high frequency slopes, these values are fixed at -1 and -3. Varying these parameters does change the confidence contours, but not by a great deal.

\vspace*{2mm}

\vspace*{2mm}

\begin{figure}
\includegraphics[width=80mm,height=90mm,angle=270]{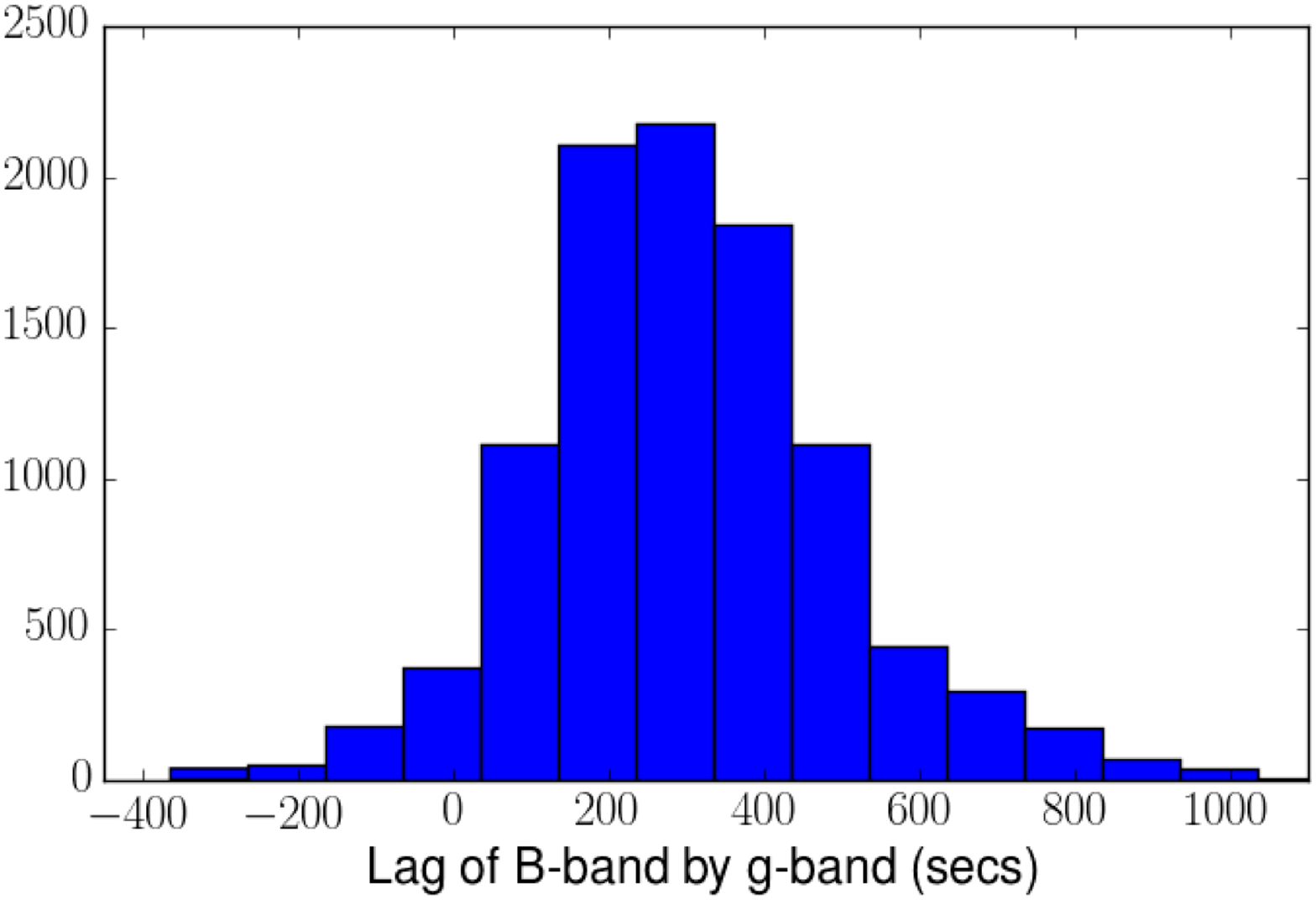}
\vspace*{-2mm}
\caption{Lag distribution from JAVELIN between the B and $g$ lightcurves from LT night 2 observations (Fig.~\protect\ref{fig:ubg_sec2}).
}
\label{fig:lt_sec2_bg_jav_lag}
\end{figure}

\begin{figure}
\includegraphics[width=80mm,height=90mm,angle=270]{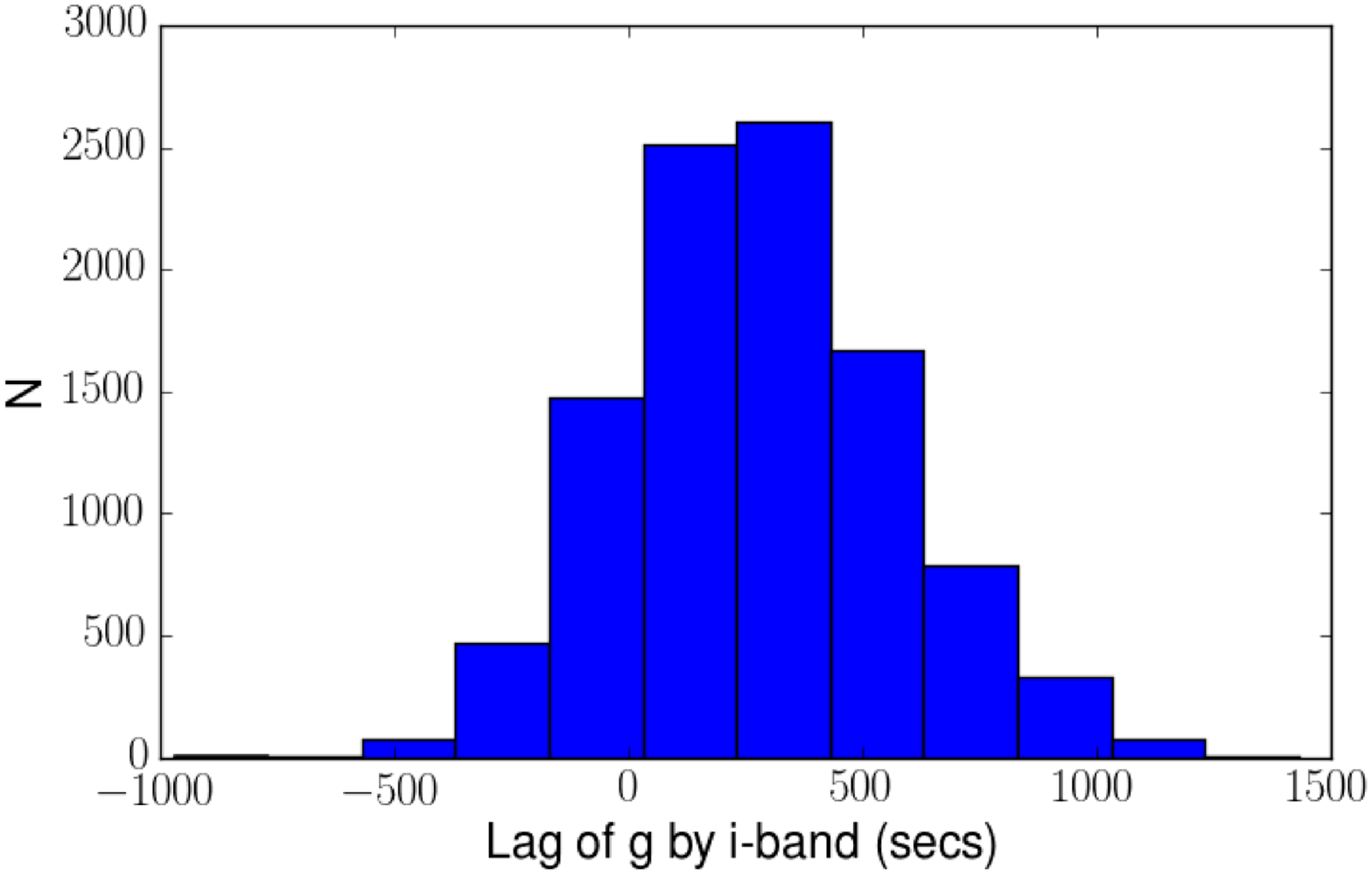}
\vspace*{-2mm}
\caption{Lag distribution from JAVELIN between the $g$ and $i$ lightcurves from LT night 3 observations (Fig.~\protect\ref{fig:griz_sec3}).
}
\label{fig:lt_sec3_gi_jav_lag}
\end{figure}

The DCF is useful for establishing the significance of a correlation but not for measuring the lag. For that purpose we use the Flux Randomisation / Random Subset Selection (FR/RSS) method of \cite{peterson98} and JAVELIN \citep{zu11_javelin,zu13_javelin}.
In Figs~\ref{fig:lt_sec2_bg_jav_lag} and \ref{fig:lt_sec3_gi_jav_lag} we show the lag distributions from JAVELIN between the B and $g$ lightcurves from night 2 and the $g$ and $i$ lightcurves from night 3 respectively. Formally the $g$ band lags the B-band by $268^{+192}_{-155}$s and the $i$ band lags the $g$ band by $267^{+309}_{-280}$s. The FR/RSS method gives similar median values, though with larger uncertainties. JAVELIN gives a lag of the $g$ band by the $r$ band of  $-74^{+294}_{-226}$s whilst FR/RSS gives $+28 \pm 436$s. For the lag of the $r$ band by the $i$ band JAVELIN gives a lag of $273^{+223}_{-365}$s and FR/RSS gives $300 \pm 500$s.
Due to the low signal to noise (S/N) of both the $u$ and $z$-band lightcurves, both FR/RSS and JAVELIN, although showing correlations, give very broad lag distributions with uncertainties $>1000$s for any correlations involving those bands. 

We conclude that the LT lightcurves show a strong correlation between all wavebands.
The lag of the $g$ band by the $B$ band is statistically robust (Fig.~\ref{fig:lt_sec2_bg_jav_lag}) and there are suggestions of wavelength-dependent lags between some of the other wavebands. However the lag uncertainties for the other bands are too large to allow accurate measurement of the lags.

\vspace*{2mm}

\vspace*{2mm}

\section{GTC HiPERCAM Observations}

\begin{figure}
\hspace*{-3mm}
\includegraphics[height=88mm,angle=270]{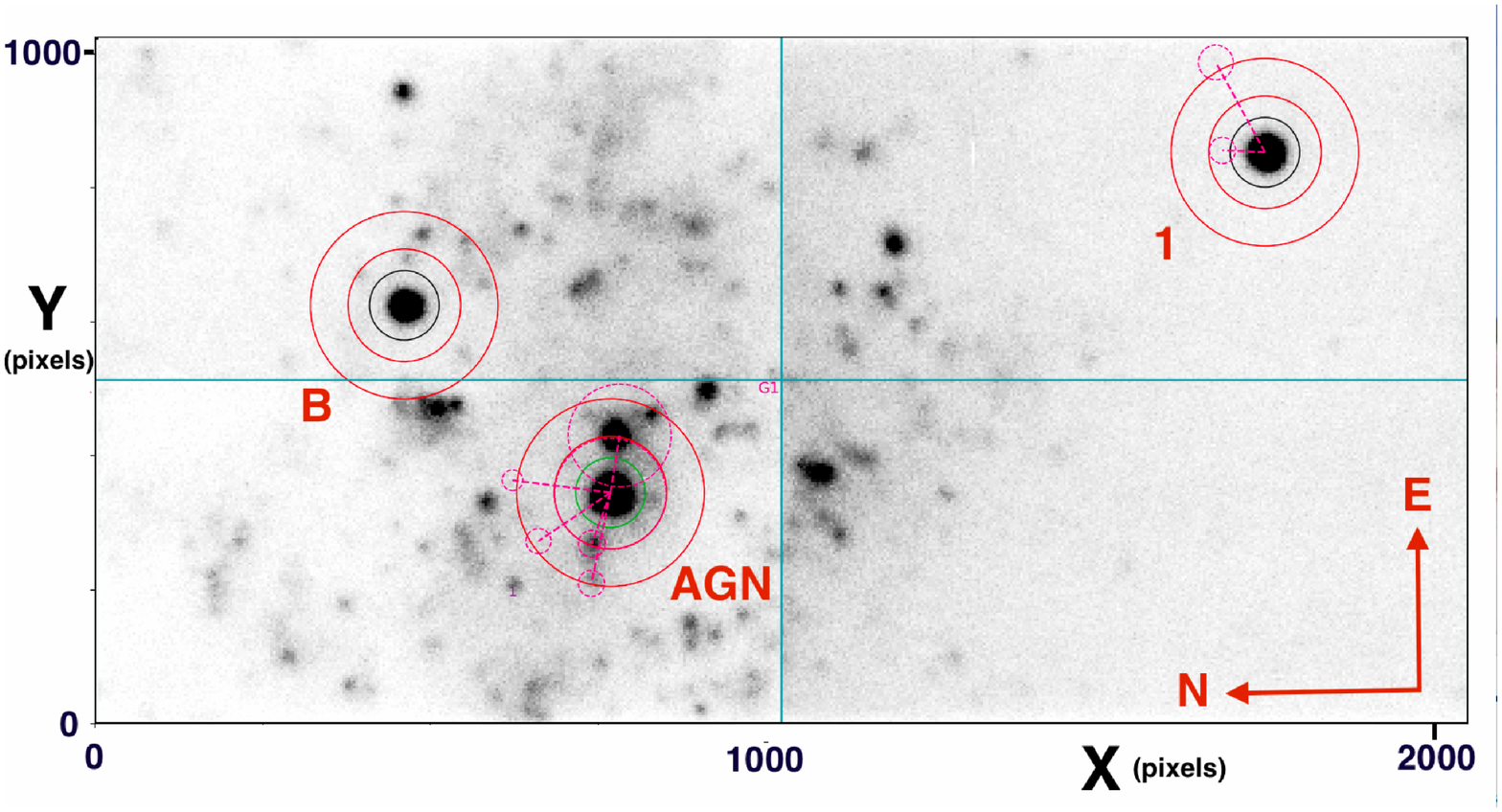}
\vspace*{-5mm}
\caption{HiPERCAM $u_{s}$-band image. The light blue lines designate the four quadrants of the CCD which are read out individually.
The AGN together with comparison stars 1 and B, also shown in Fig.~\protect\ref{fig:lt_finder}, are labelled. The inner circles, green around the AGN and black around the comparison stars, of radii 52 pixels (4.21 arcsec), define the object apertures. The sky background lies between the two surrounding solid red circles (radii 84 and 140 pixels, i.e. 6.80 and 11.34 arcsec respectively). The dashed red circles indicate stars in the background regions which have been masked out. The dashed straight lines simply connect the masked areas to the central object that they are associated with. For the relatively bright star to the east of the AGN, a large masking radius is chosen so as to encompass also a nearby fainter star. However only the part of the image which is also within the background annulus is masked. The object aperture is unaffected. Note that, opposite to the conventional projection used in Fig.~\protect\ref{fig:lt_finder}, east is to the right of north. The HiPERCAM major axis was tilted very slightly east of north. 
}
\label{fig:hipercam_finder}
\end{figure}

\subsection{Observations}
Following the proof of concept LT observations, 
NGC~4395 was observed for just over 6 hours from 2018-04-16 20:58:14.665 UTC (MJD 58224.8737808) to 2018-04-17 03:06:07.266 UTC (MJD 58225.1292507) with HiPERCAM
on the 10.4m Gran Telescopio Canarias (GTC)\footnote{\url{http://www.gtc.iac.es/}}.

HiPERCAM \citep{dhillon21} is a quintuple-beam high-speed astronomical imager. Incoming white light is split into 5 colours, each sent to a separate camera, with each camera  containing a 2048$\times$1024 pixel frame-transfer CCD which can read out at up to 1000 frames per second. The 5 filters have bandpasses which very closely resemble the standard Sloan {\it ugriz} bands but have improved throughput, particularly in the \us\, band where throughput is 41 per cent higher. These Super SDSS filters\footnote{\url{http://www.vikdhillon.staff.shef.ac.uk/ultracam/filters/}}
are referred to as $u_{s}$, $g_{s}$, $r_{s}$, $i_{s}$, and $z_{s}$.
The plate scale is 0.081 arc seconds per pixel giving a field of view of 2.8$\times$1.4 arcminutes. The integration time was 3s for the  $g_{s}$, $r_{s}$, $i_{s}$, and $z_{s}$
bands and 15s for $u_{s}$. 

We arranged the field of view to encompass the two comparison stars, 1 and B, which are shown in Fig.~\ref{fig:lt_finder}. The data were reduced using the HiPERCAM software suite\footnote{\url{http://deneb.astro.warwick.ac.uk/phsaap/hipercam/docs/html/}}.
A HiPERCAM $u_{s}$ image
is shown in Fig.~\ref{fig:hipercam_finder}. Object (green or black) and sky apertures (solid red)
are shown around all objects. Masked regions, to remove stars from background areas (dashed red lines) are also shown. Details are given in the figure caption. Note that this image is mirrored around the north-south axis so east is to the right of north, not the usual left. 
The software then produces count rates in the various apertures, X and Y coordinates and FWHM of targets for all of the separate images thus allowing the production of lightcurves.

\begin{figure}
\hspace*{-13mm}
\includegraphics[width=60mm,height=90mm,angle=270]{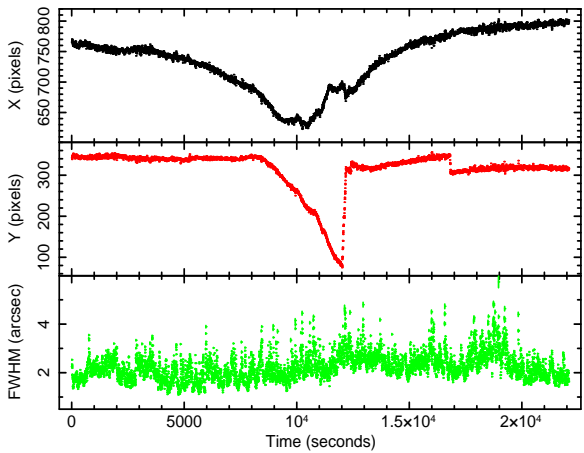}
\caption{Tracking and seeing plots of the HiPERCAM data.}
\label{fig:xy}
\end{figure}

During the observations the sky was dark and photometric, but the seeing was quite variable, ranging between $\sim 1.5$ and 4 arcseconds. The seeing, defined as the FWHM of the unresolved core of the AGN, together with the X and Y CCD pixel coordinates of the AGN, are shown in Fig~\ref{fig:xy}. We use the FWHM of the AGN rather than that of star 1 simply for computational convenience but the values are almost exactly the same.
In the g-band discussed here the ratio of the FWHM of the AGN to that of the star is constant at $1.02 \times$ with a variance of $8.4 \times 10^{-5}$ so any variations are tiny compared to the factors of a few in the seeing itself. 
The field of view remained as shown in Fig.~\ref{fig:hipercam_finder} until the source approached transit, which occured close to the middle of the observation.
At this time the field passed very close to the zenith where tracking is not so precise and, at the time of these observations, there was no autoguider at the HiPERCAM focus of the GTC. Hereafter we refer to times after the start of the observation.
After 8000s, the field of view drifted slowly to the west, as shown by the large change in the Y-coordinate of the AGN (Fig.~\ref{fig:xy}). 
After 12000s the field of view was manually adjusted back to the original position. During these movements, star B crossed into the CCD quadrant containing the AGN and was then moved back. However star 1 remained within its original CCD quadrant at all times. Additionally, as the background surrounding star 1 is empty, unlike that surrounding star B, star 1 is preferred as the comparison star. 
There are a number of small bumps in the lightcurves between 10000 and 14000s, most of which can be associated with rapid changes in field location and so are not considered to be physically real. Data between these two times is therefore not considered to be reliable and so is not considered in this analysis. 

\begin{figure}
\hspace*{-13mm}
\includegraphics[width=60mm,height=90mm,angle=270]{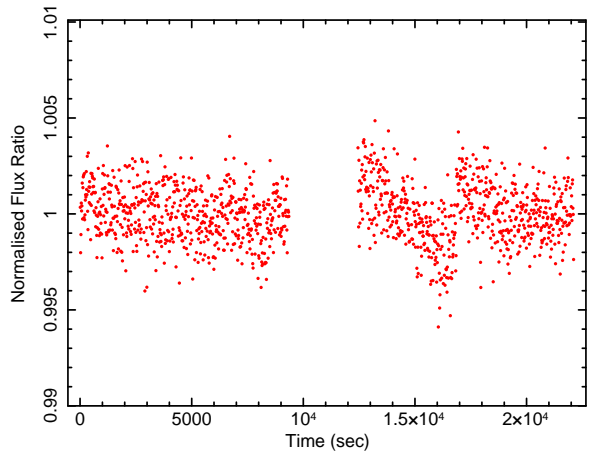}
\caption{Normalised ratio of the $g_{S}$-band count rates of stars 1 and B.}
\label{fig:star_ratio}
\end{figure}

\begin{figure*}
\includegraphics[width=120mm,height=180mm,angle=270]{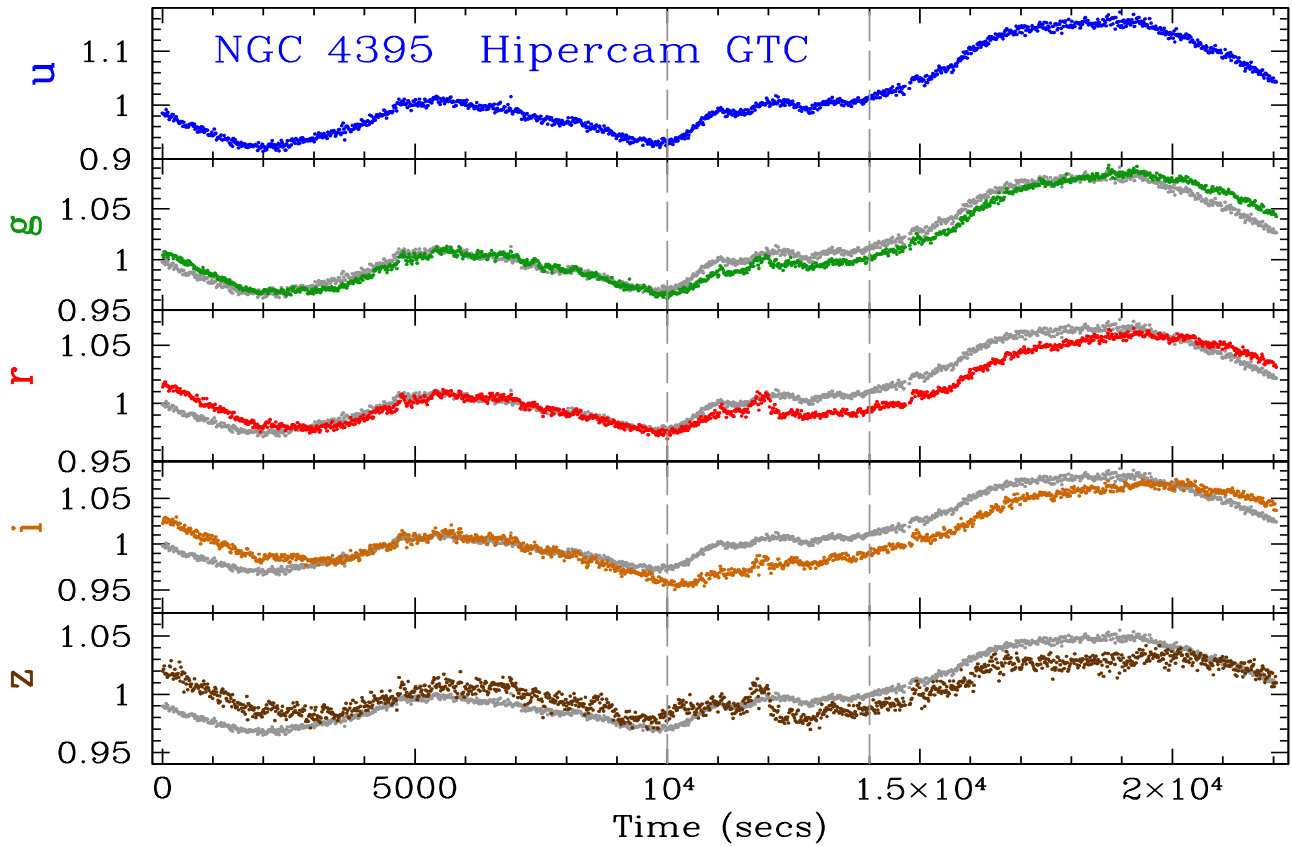}
\vspace*{-2mm}
\caption{
Observed $u_{s}, g_{s}, r_{s}, i_{s}$ and $z_{s}$ lightcurves, normalised to the mean, for NGC~4395 from HiPERCAM on the GTC binned to 15s exposures. Statistical errorbars are too small to be seen and are, from \us\, to \zs\, respectively, 0.0047, 0.00165, 0.00175, 0.00243 and 0.00356. The lightcurves between 10,000 and 14,000s, as indicated by the dashed grey lines, are affected by loss of tracking and motion of the target and comparison stars over the CCDs, as described in the text. For each of the \gs\, to \zs\, bands, the \us\, lightcurve, with variability amplitude scaled to be the same as for the appropriate \gs\, to \zs\, band, is underlaid in grey, for comparison. 
}
\label{fig:hipercam_lcs}
\end{figure*}

At 14650s an unplanned instrument rotator unwrap occured. As it is not possible to follow the motion of objects through that rotation, the lightcurve was produced in two parts, before and after that time. There is a simultaneous small ripple in all lightcurves at that time.
Between 16780 and 16840s there was a small change in the Y-positions but this change occured sufficiently slowly that we were able to accurately follow the movement so the photometry was not affected.

In Fig.~\ref{fig:star_ratio} we show the normalised ratio of the count rates between star 1 and star B. This figure only covers the time when star B was in the CCD quadrant in which it started. There is a gap in the middle when star B was in a different CCD quadrant, i.e. the one in which the AGN is always located. During the first part of the observation when the AGN and both stars remained on their original CCD quadrants, there is no change in the ratio of star 1 to star B. Formally the error-subtracted excess variance of this part of the lightcurve is $-6.68 \times 10^{-8}$ and the average standard deviation is $1.37 \times 10^{-3}$. There is no evidence for variability in either comparison star on the timescales of the observation and so, in principle, either star may be used to measure changes in AGN intensity. We do, however, note a small change in the stellar ratio after the gap shown in Fig.~\ref{fig:star_ratio} and around 16800s, corresponding to a sudden change in position of the objects. As star 1 remains on its original CCD quadrant throughout the observation and also has a lower surrounding background and so is less likely to suffer from spurious position-induced changes in intensity, we use star 1 to determine changes in the AGN. The resultant lightcurves are show in Fig.~\ref{fig:hipercam_lcs}.

\subsubsection{Seeing Effects}
\begin{figure}
\hspace*{-8mm}
\includegraphics[width=70mm,height=100mm,angle=270]{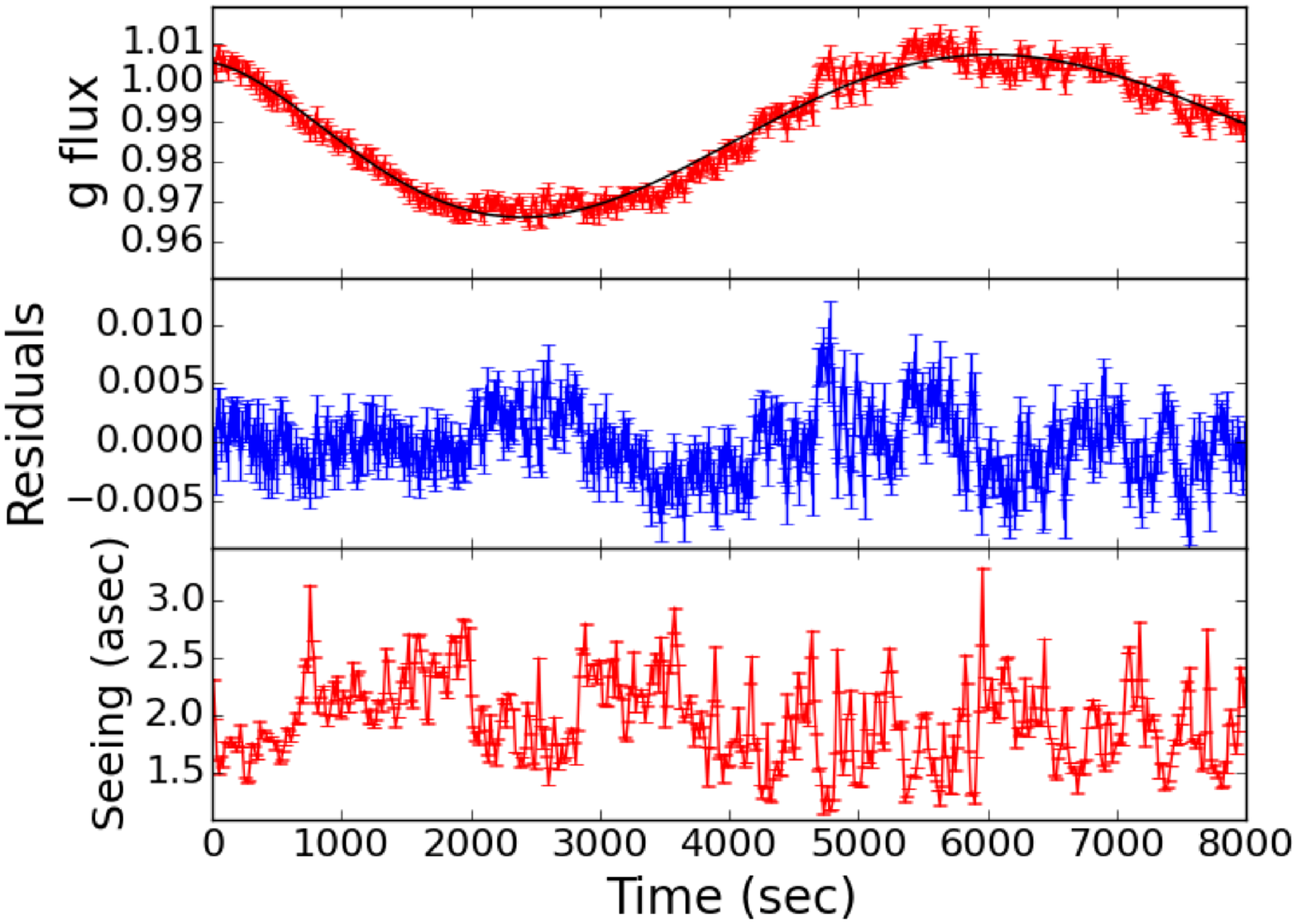}
\caption{{\it Top panel:} $g_{s}$-band normalised flux (red errorbars) with 6th order polynomial fit (black) for the first 8000s of the GTC HiPERCAM observation. {\it Middle panel:} Flux residuals relative to the polynomial fit. {\it Bottom panel:} Seeing, i.e. FWHM of the AGN point spread function (PSF). An anti-correlation with the residuals is apparent.}
\label{fig:seeing8000}
\end{figure}

Although the lightcurves are generally quite smooth, there are occasional changes of flux on timescales of a $\sim$few hundred seconds. As the seeing can vary on timescales at least as short we investigated whether seeing changes might cause flux changes which might affect lag measurement.
Assuming that the slower changes represent real physical changes in the luminosity of the AGN, we fitted 6th order polynomials to the lightcurves and measured the residuals. An example fit to the \gs\, lightcurve, with residuals, and seeing, is shown in  Fig~\ref{fig:seeing8000}. Whilst recognising that this fit does make assumptions about what a real physical variation in the AGN is, a lower-order polynomial does not fit the long-timescale variations as well and a higher order does not improve the fit. We therefore take it as a first attempt at determining the effect of seeing on flux. 

\begin{figure}
\hspace*{-8mm}
\includegraphics[width=60mm,height=100mm,angle=270]{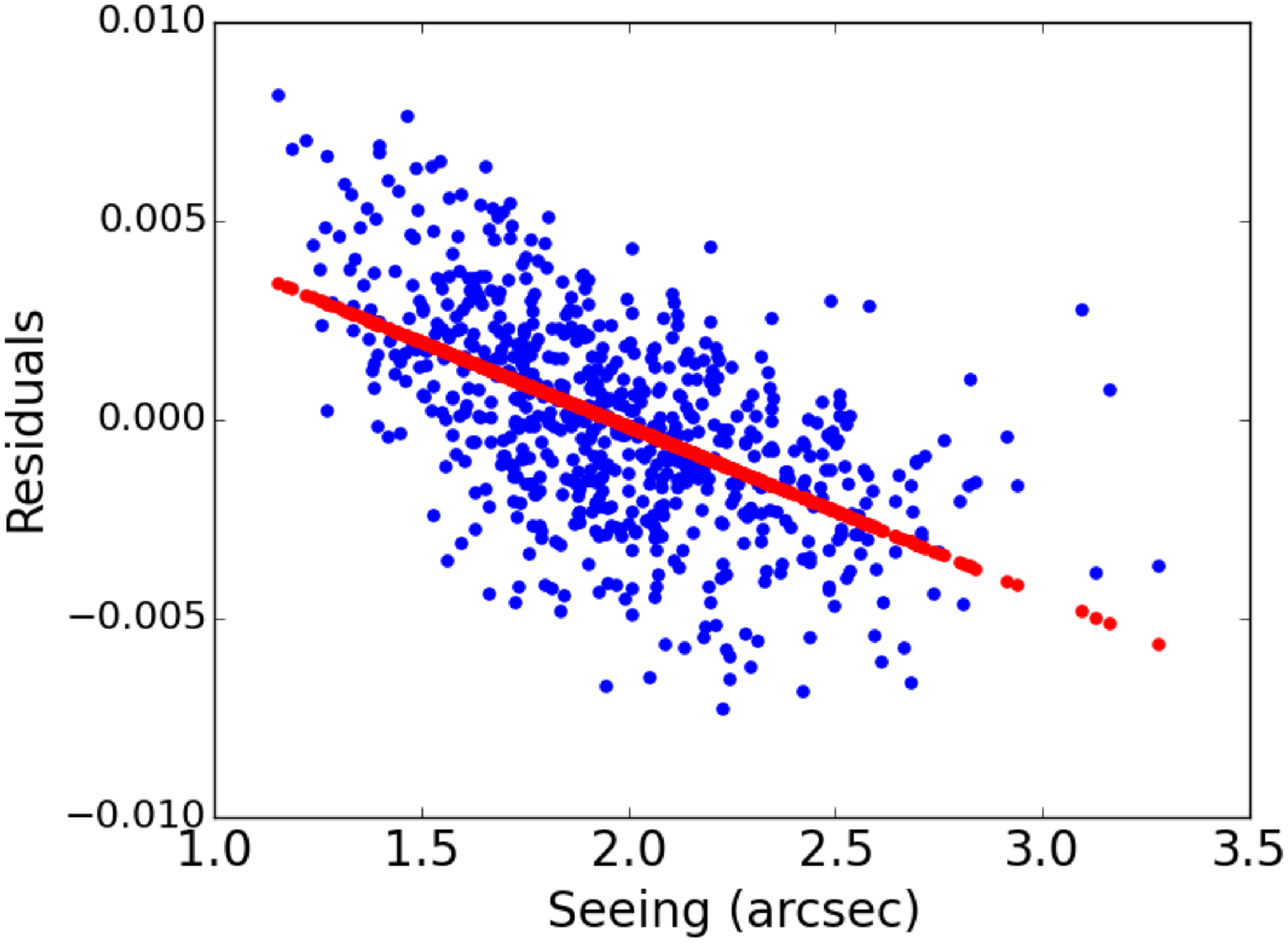}
\caption{Residuals from Fig~\ref{fig:seeing8000} vs seeing.}
\label{fig:resid_see}
\end{figure}

\begin{figure}
\hspace*{-8mm}
\includegraphics[width=60mm,height=100mm,angle=270]{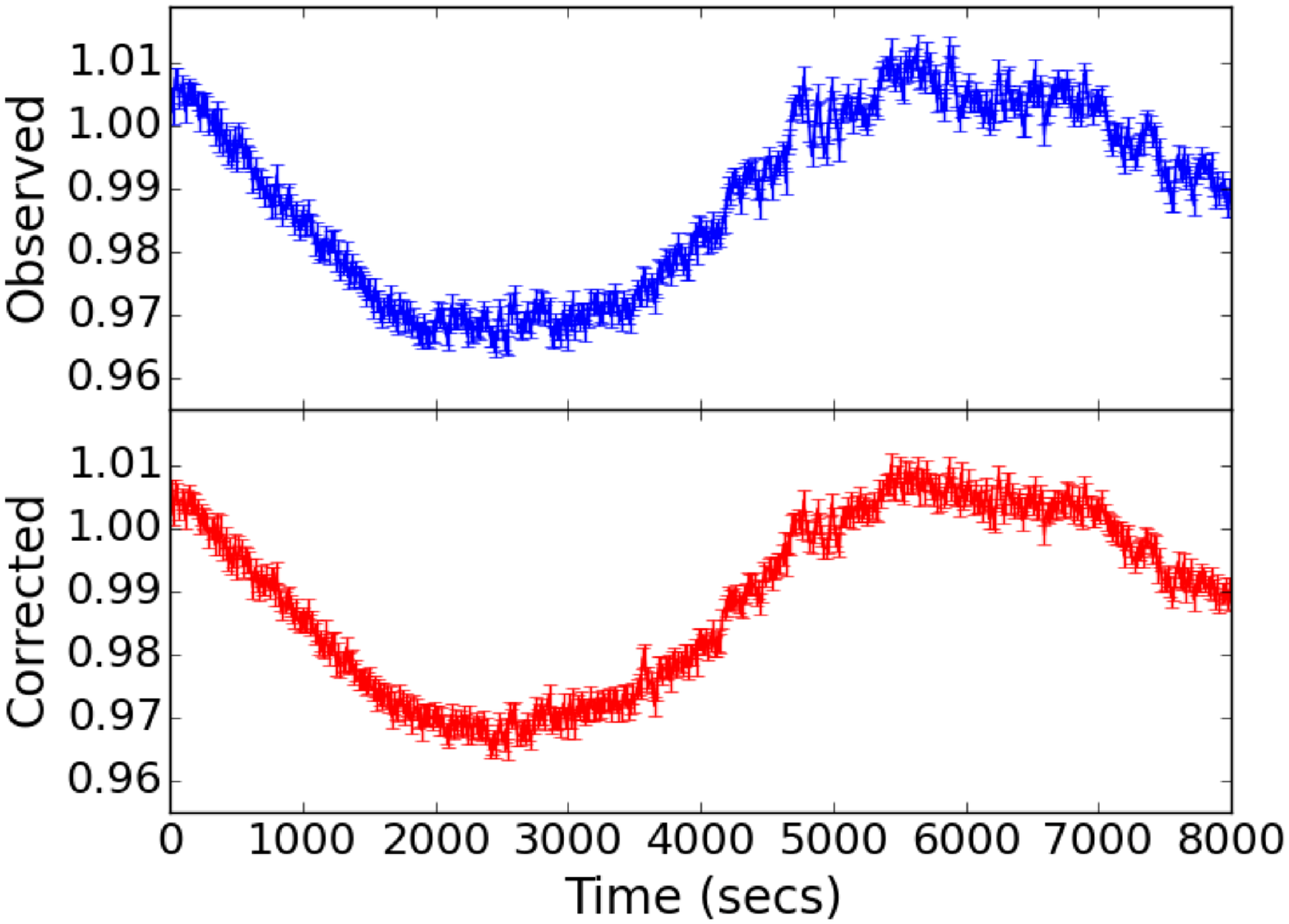}
\caption{{\it Top Panel} The first 8000s of the \gs-band lightcurve. {\it Bottom Panel}. Corrected \gs-band lightcurve, using the correction shown in Fig~\ref{fig:resid_see}
}
\label{fig:observed_corrected}
\end{figure}

In Fig.~\ref{fig:resid_see} we plot the residuals against seeing.
Although there is considerable scatter, there is a definite trend for the residuals to become larger as the seeing improves, i.e. the residuals and seeing are anti-correlated. Changes with seeing in the fraction of light in the object apertures should be the same for the AGN core and for the comparison star. Therefore this anti-correlation may be more related to changes in the light in the background apertures and may not necessarily be repeatable in imaging of other objects in different environments.
We calculate the best fit to that relationship and apply the resultant correction to the observed lightcurve, producing a corrected lightcurve which is shown in Fig.~\ref{fig:observed_corrected}. The amplitudes of small-timescale variations are lower in the corrected lightcurves. In the next section we consider the effect of this correction on lag measurements.

\vspace*{-2mm}
\vspace*{-2mm}

\subsection{Fractional Variability}

We list, in Table~\ref{tab:fvar}, the percentage fractional variability ($F_{\rm var}$) in the different bands as we are aware that these numbers may well be of interest to other observers in other contexts. The numbers in Table~\ref{tab:fvar} are not corrected for seeing. Correction for seeing reduces these numbers by only 4 per cent.
We list the values as derived for the first 8000s and for the complete lightcurves with the section from 10,000 to 14,000s removed. The \us~ band is approximately twice as variable as the \gs~ band but the decrease in $F_{\rm var}$ with increasing wavelength is not so pronounced at longer wavelengths. The \is~ band is, in fact, slightly more variable than both the \rs~ and \zs~ bands. These observations are consistent with the standard paradigm of the longer wavelengths coming from a larger area of the disc, more distant from the black hole.
As $F_{\rm var}$ is larger on longer timescales, we conclude that these observations are sampling the high frequency part of the optical variability power spectrum where the power spectrum is steep. The power spectrum of these observations will be discussed elsewhere (Beard et al, in prep).

We also list, in Table~\ref{tab:fvar} the magnitudes of Star 1 in the various bands as taken from SDSS. The SDSS coordinates are 12 25 51.28 +33 31 26.76 (J2000). We also list
the count rate ratios of the AGN to Star 1 from the first 8000s of the observations and hence the derived magnitudes of the AGN. These numbers should be treated with caution and treated as indicative rather than precise as we did not do detailed multiband photometric calibration of the HiPERCAM observations.

\begin{figure}
\hspace*{-2mm}
\includegraphics[width=80mm,height=85mm,angle=270]{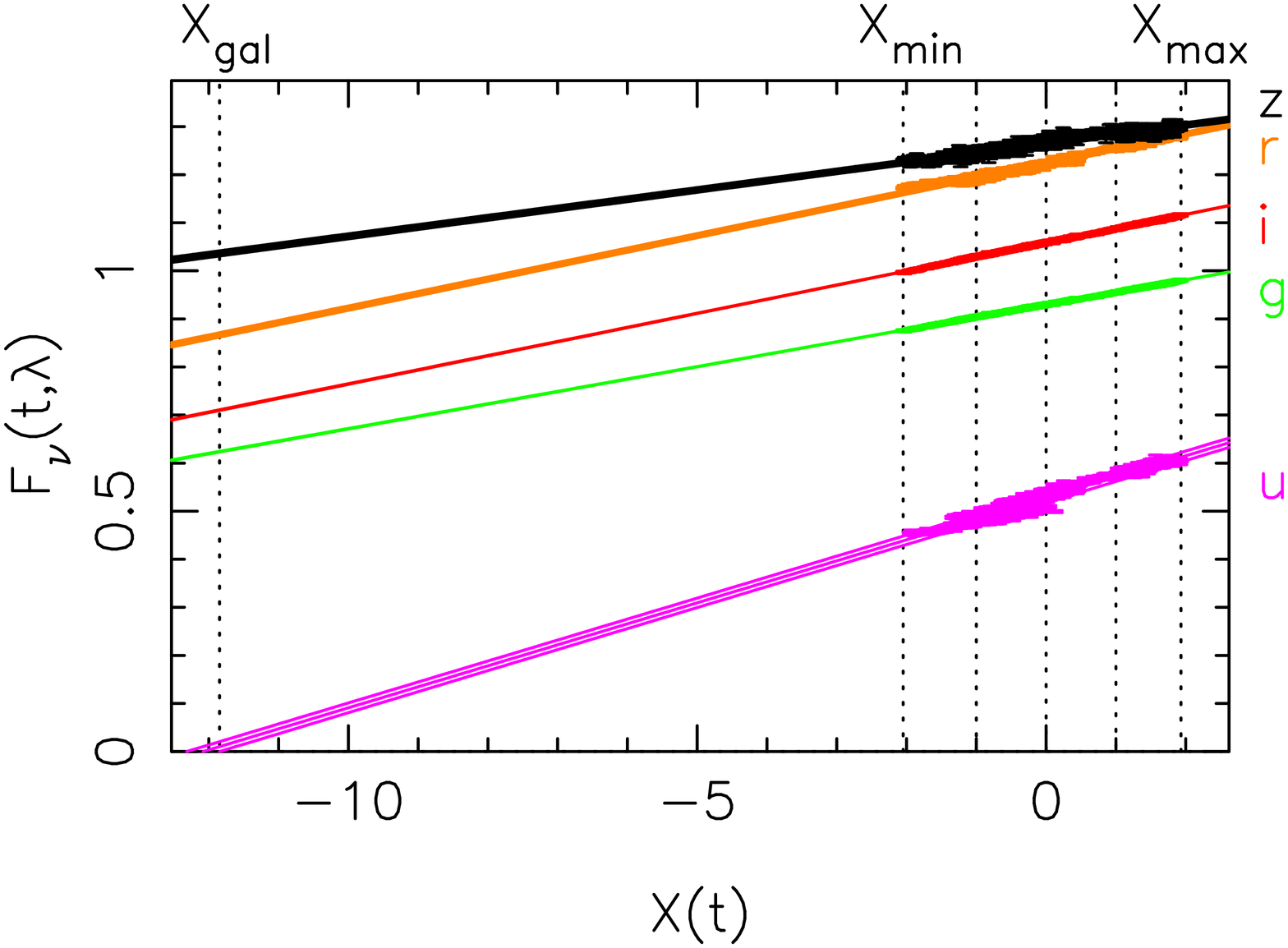}
\caption{$F_{\nu}(t, \lambda)$ 
versus an assumed underlying driving lightcurve, X(t), as defined in equation (1), for each of the HiPERCAM bands.}
\label{fig:fx}
\end{figure}

\begin{figure}
\hspace*{-2mm}
\includegraphics[width=80mm,height=85mm,angle=270]{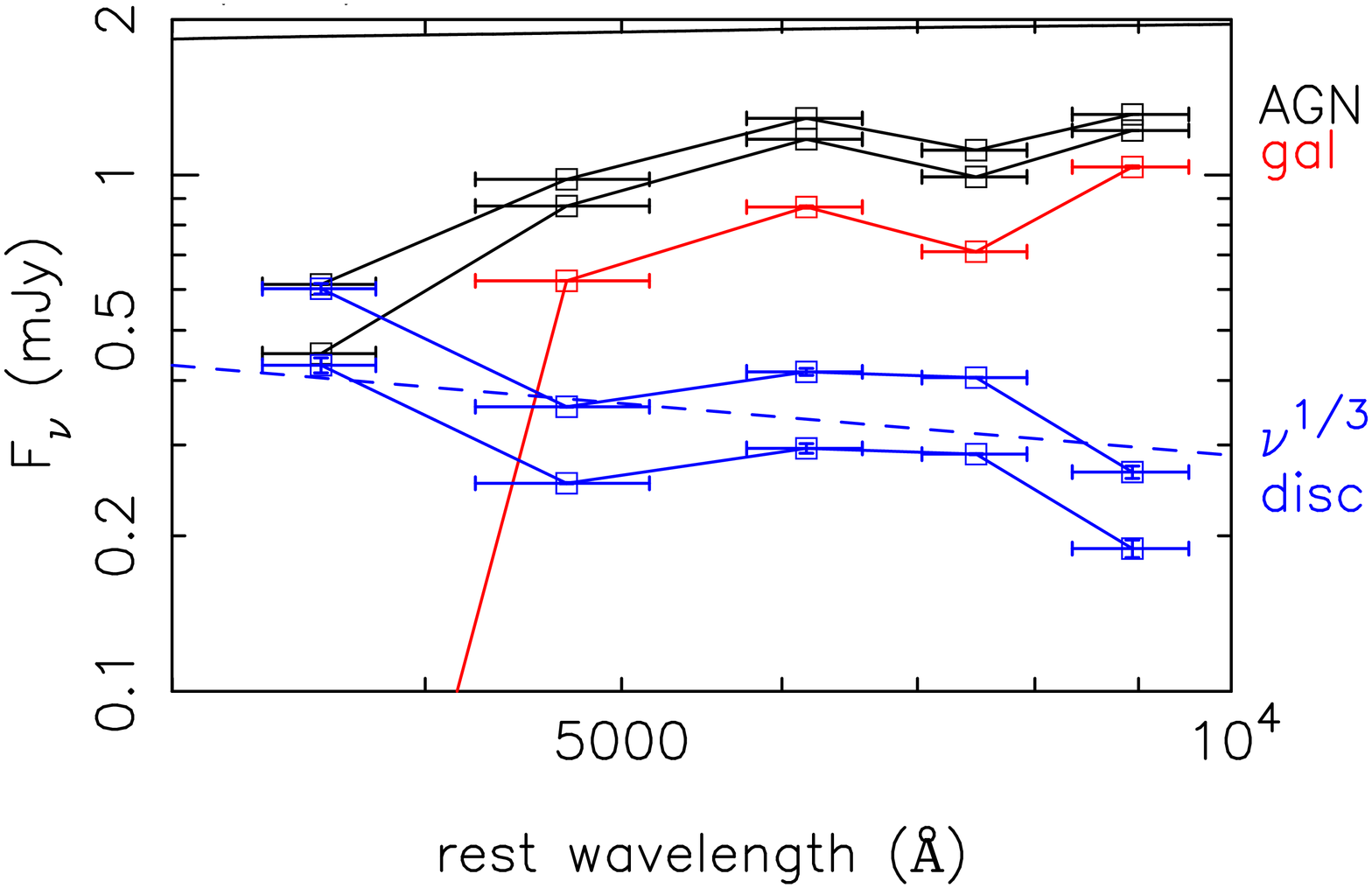}
\caption{The two black lines give the total fluxes in the HiPERCAM bands for the minimum and maximum fluxes observed as defined at $X_{min}$ and $X_{max}$ in Fig.~\ref{fig:fx}. The red line is the constant galaxy contribution defined at $X_{gal}$ in Fig.~\ref{fig:fx} where the \us-band flux extrapolates to zero. The two blue lines are the variable component, defined by the fluxes in the  HiPERCAM bands at $X_{min}$ and $X_{max}$. The dashed blue line represents $F_{\nu} \propto \nu^{1/3}$ which applies to simple accretion discs.
}
\label{fig:seds}
\end{figure}

We have carried out flux/flux analysis similar to that described in \cite{mch18}.
We fit the fluxes in mJy, i.e. $F_{\nu}$, 
as a function of time with a linear model 
\[ F_{\nu}(t, \lambda) = A_{\nu}(\lambda) + S_{\nu}(\lambda)~ X(t) \,\,\,\,(1) \]
$X(t)$ is the underlying lightcurve, which is the same for all bands. It is normalised to a mean 0 and rms 1. $A_{\nu}(\lambda)$ is the mean spectrum and $S_{\nu}(\lambda)$ is the rms spectrum for each band. 
The results are shown in Figs.~\ref{fig:fx} and ~\ref{fig:seds}. 

In Fig.~\ref{fig:fx} we see linear relationships between the fluxes in all bands and $X(t)$ which tells us that the simple model of a combination of a constant and a variable component is a good representation of the data. The lack of curvature tells us that the SED of the variable component does not change with luminosity. 

From the intercepts of the flux/flux plots we can disentangle the spectra of the constant and variable components, as described in the caption to Fig.~\ref{fig:seds}. The constant (galaxy) component is red and the variable one (disc) is blue. 
The apparent luminosity-related colour changes are a result of combining different fractions of these two components.


\begin{table}
\begin{tabular}{llllll}
Band & $F_{\rm var}$ (\%) & $F_{\rm var}$ (\%) & Ratio & Star 1 & AGN \\
     & First & All bar                      & AGN/Star1& Mag& Mag\\ 
     & 8000s & 10-14ks &&&  \\
&& \\
{\it $u_{s}$}& 2.92 & 7.71 & 1.1482    &    17.41 &    17.26 \\
{\it $g_{s}$}& 1.44 & 4.01 & 0.73566   &   16.23  &  16.56 \\
{\it $r_{s}$}& 1.09 & 2.74 & 0.6153    &    15.71  &   16.24\\
{\it $i_{s}$}& 1.17 & 3.05 & 0.4666    &    15.56  &  16.39\\
{\it $z_{s}$}& 1.04 & 1.71 & 0.5108    &    15.45  &   16.18 \\ 
\end{tabular}
\caption{Per centage fractional variability of the HiPERCAM observations as a function of both waveband and timescale together with the SDSS magnitude of Star 1, the count rate ratio of the AGN compared to Star 1 as derived from the first 8000s of the observations and the derived magnitude of the AGN.
}
\label{tab:fvar}
\end{table}

\subsection{Interband Lags}
\label{sec:lags}

\begin{figure*}
\hspace*{-5mm}
\includegraphics[width=60mm,height=75mm,angle=90]{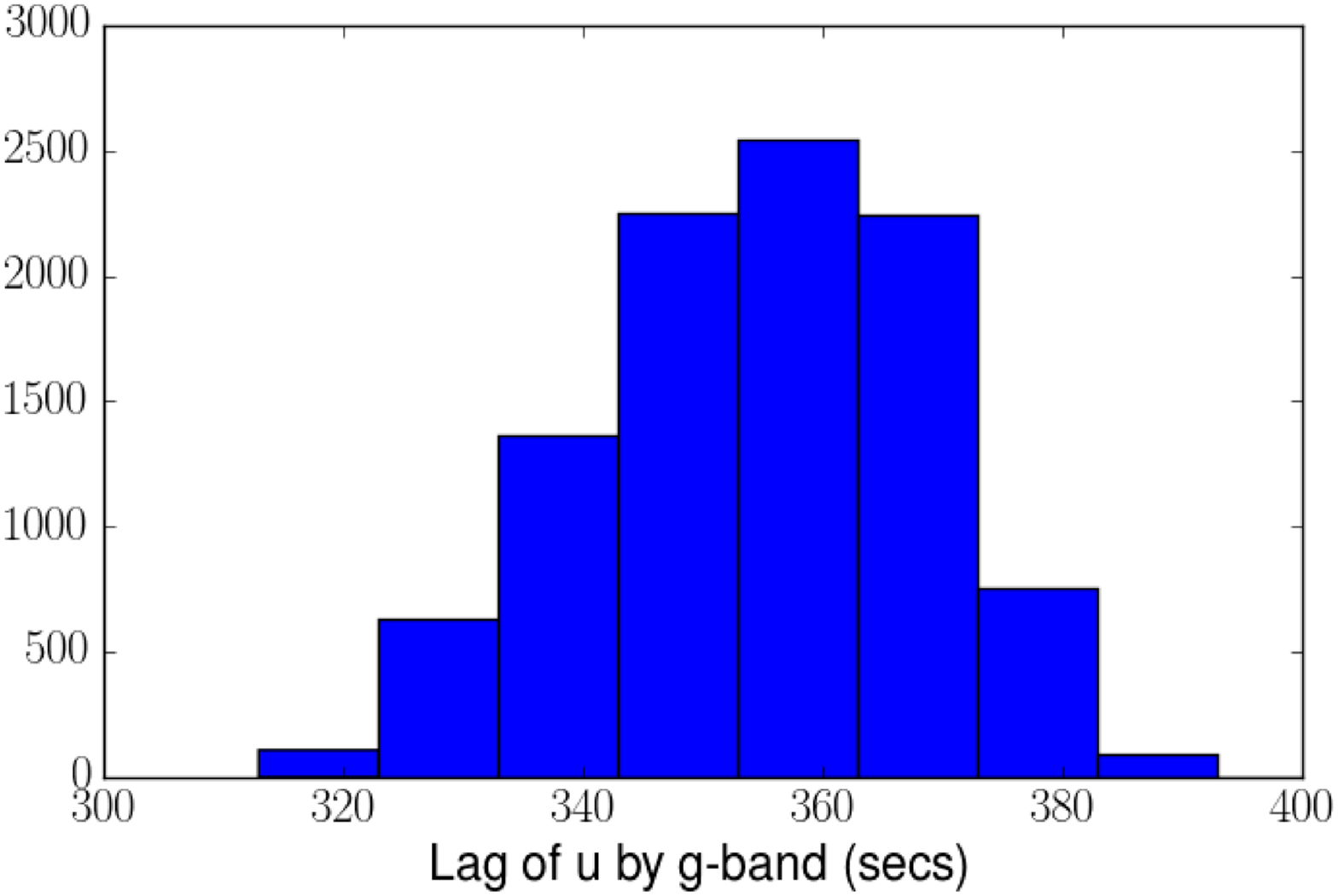}
\hspace*{5mm}
\includegraphics[width=60mm,height=75mm,angle=90]{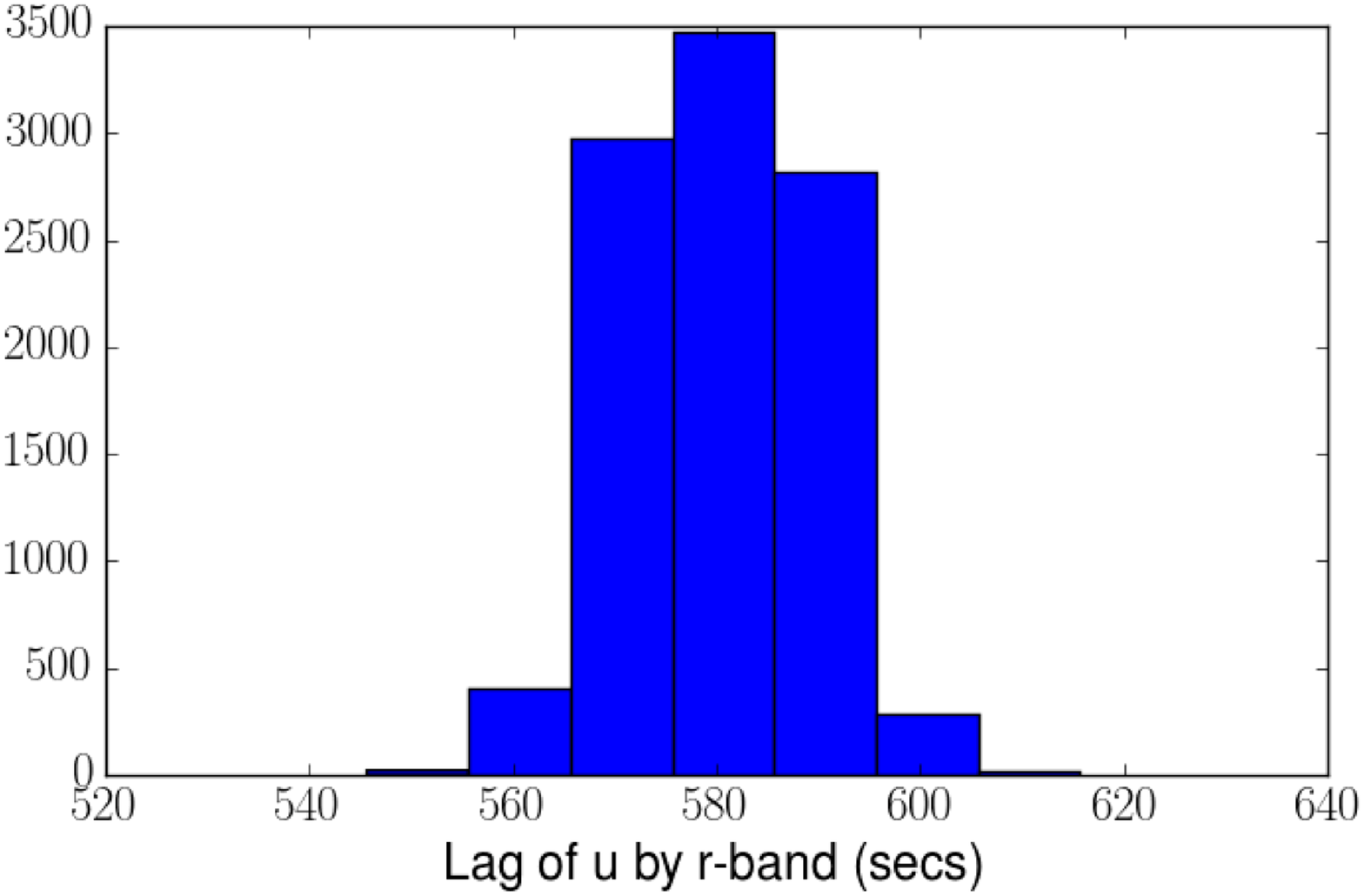}
\caption{HiPERCAM lag probability distributions from {\sc JAVELIN} for {\it (Left)} the \gs, and {\it (Right)} the \rs,  bands relative to the \us\, band. The distributions are for the first 8000s of the observation, with lightcurves corrected for seeing changes.
}
\label{fig:javlagdistributions_gr}
\end{figure*}

\begin{figure*}
\hspace*{-5mm}
\includegraphics[width=60mm,height=75mm,angle=90]{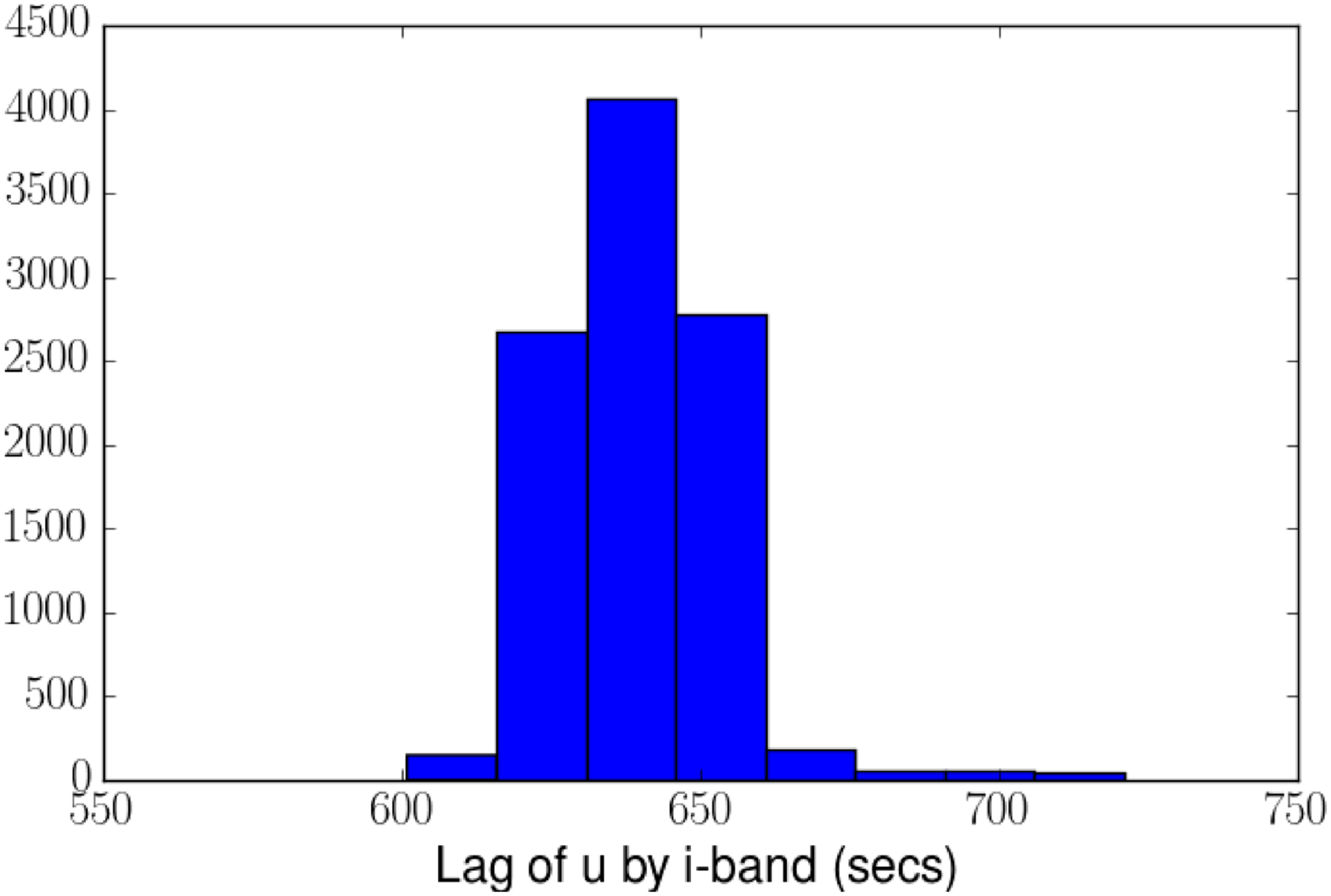}
\hspace*{5mm}
\includegraphics[width=60mm,height=75mm,angle=90]{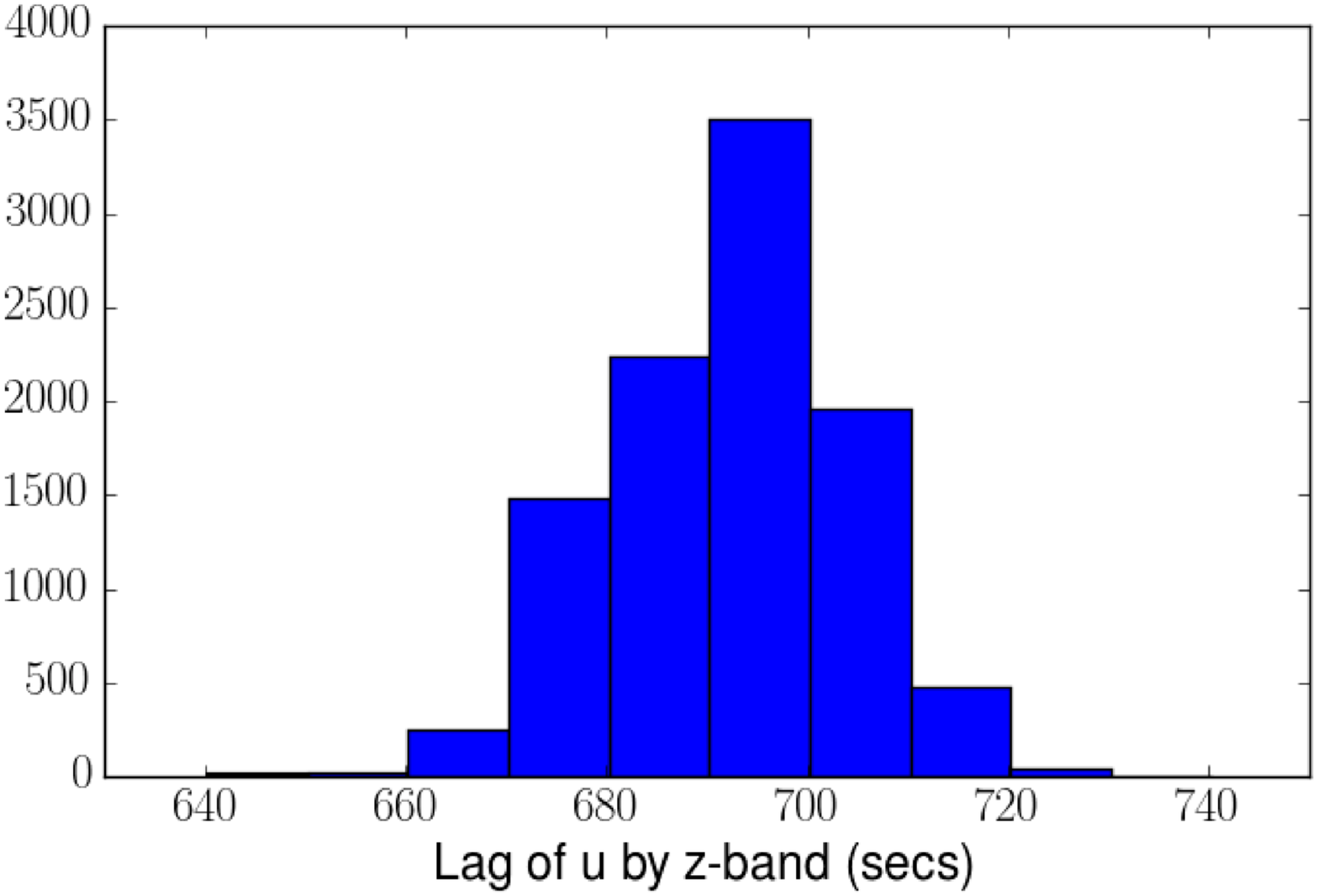}
\caption{As for Fig.~\ref{fig:javlagdistributions_gr} except for  {\it (Left)} the \is\, and  {\it (Right)} the \zs\, bands.
}
\label{fig:javlagdistributions_iz}
\end{figure*}

Given the concerns regarding possible spurious count-rate variations induced by rapid positional changes, we initially restrict our lag measurement to the first 8000s of the observation which are largely unaffected.
In Table~\ref{tab:lags} we give the lags, relative to the \us\, band, for the other HiPERCAM bands for the first 8000s of the observations. We use lightcurves with 15s sampling as that is the minimum available for the \us\, band.
We list the lags as measured using JAVELIN \citep{zu11_javelin,zu13_javelin} and using the centroid lags as calculated using the FR/RSS correlation function method \citep{peterson98}. Here we apply both flux randomisation and random subset selection. Thus the errors here are larger than would be obtained if we chose only one of the methods. In Figs.~\ref{fig:javlagdistributions_gr} and ~\ref{fig:javlagdistributions_iz} we show the lag distributions from JAVELIN.
We list the lags as measured both on the uncorrected data and on data corrected for seeing effects. There is no significant difference between any of the methods except that the uncertanties for the FR/RSS method are larger as, in the RSS method, only 70 per cent of the available data are used and gaps are introduced into the lightcurves. Thus the associated uncertainties are very conservative. However, compared to all previous AGN lag measurements, as presented in the many RXTE and Swift-based papers listed in the first paragraph of the Introduction here, the errors are still very small.

We have also calculated the lags for the full lightcurve, excepting the middle region from 10,000 to 14,000s where the tracking was particularly poor. Although these data are therefore not of quite as good quality as that for the first 8000s, they provide a preliminary check on whether the lags might depend on which small section of the lightcurves one chooses, which might be affected by some small non-astronomical variation which we have not properly accounted for. We have not attempted to correct this full lightcurve for seeing variations. The only possible difference between the first 8000s and the full lightcurve is that the average luminosity level is lower in the first 8000s than in the full lightcurve. The measured lags, using JAVELIN, relative to \us\, are 
$415^{+13}_{-4}, 622^{+13}_{-14}, 689^{+30}_{-45}$  and $669^{+9}_{-5}$s
respectively for the \gs, \rs, \is\, and \zs\, bands. Although the lags of the \gs\, band and to a lesser extent also the \rs\, band are a little longer than in the first 8000s of the observations (Table~\ref{tab:lags}), the lags of \is\, and \zs\, are quite similar. These small differences are in agreement with what we expect from a higher illuminating luminosity. 

There are some short-timescale bumps in the lightcurves which remain even after correction for seeing variations. For example there is a bump at around 4800s which is quasi-simultaneous in all bands. It is possible that the seeing correction model has not been sophisticated enough as the bump does correspond to a decrease in seeing. We have therefore measured lags excluding the section from 4500 to 8000s but it does not make a noticeable difference to the lags. Overall we conclude that the lags listed in Table~\ref{tab:lags} are a fair reflection of the true lags.

In Table~\ref{tab:lags} we also give the effective wavelength of each band for these particular filters as used on HiPERCAM on the GTC and taking account of atmospheric transmission. We derived these values, which are the `Bandpass Average Wavelengths', using the {\sc pysynphot} package\footnote{\url{https://github.com/StuartLittlefair/ucam_thruput}} which is specifically designed for HiPERCAM. We note that the values derived here are actually very similar to the central wavelengths listed on the HiPERCAM filter web site\footnote{\url{http://www.vikdhillon.staff.shef.ac.uk/ultracam/filters/}},
i.e. 3526, 4732,  6199, 7711 and 9156\AA~ respectively for \us, \gs, \rs, \is\, and \zs\, for which slightly simpler assumptions are made. Here we use the effective, or Bandpass Average Wavelengths, given in Table~\ref{tab:lags}.

\subsection{Referencing the HiPERCAM lags to the X-ray band}
\label{referencing}

The HiPERCAM lags are all measured relative to the \us\, band but to obtain full lag-wavelength coverage it is necessary to reference the HiPERCAM lags to the X-ray band. There has been a previous measurement with XMM-Newton of the lag of the UVW1 band (2910 \AA), observed using the Optical Monitor in fast mode, relative to the EPIC X-ray 0.5-10 keV band, of $473^{+47}_{-98}$s \cite{mch16}. 
The X-ray and UVW1 lightcurves overlap almost completely and are both of good quality, resulting in a correlation exceeding 99 per cent significance. The X-ray and UVW1 lightcurves are still available so we have repeated the lag analysis. We confirm the original result, measuring a lag of $430^{+50}_{-80}$s. Although it is of lower statistical significance, this result is quite consistent with the measurement of a $\sim400$s lag of the UVW2 band (1920 \AA) relative to the X-rays \citep{cameron12}.

Ground-based $g$-band observations accompany most of the X-ray and UVW1 lightcurves from which a $g$-band to X-ray lag of $788^{+44}_{-54}$s was derived. However the $g$ band lightcurve had to be made by adding together lightcurves from 6 separate ground-based observatories, all with different S/N, and mostly of lower S/N than either the X-ray or UVW1 lightcurves so the confidence level of the correlation barely reached 90 per cent.  We also note that although the JAVELIN lag distribution for the X-ray to UVW1 is symmetric, with no long tails, the X-ray to $g$-band lag has a long tail up to longer lags ($\sim900$s). Unfortunately the $g$-band lightcurve can no longer be found so we cannot repeat this analysis.
We therefore consider the X-ray to $g$-band lag to be less reliable than the X-ray to UVW1 lag and therefore we reference the HiPERCAM lags to the UVW1 lag.

The effective wavelength of the UVW1 filter on the XMM-Newton Optical Monitor, including the entire in-flight telescope response, is 2910 \AA. This effective wavelength is not very different from that of the HiPERCAM \us\, band (3694 \AA), but is different enough that we should estimate the small offset expected in the lags. In any reasonable scenario, i.e. whether the UV/optical variability is dominated by reprocessing from a disc or from the BLR, we expect this lag to be positive.  Scaling the lag to the 4/3 power of wavelength \citep{cackett07}, as expected from reprocessing by a disc with the temperature profile defined by \cite{shakura73}, would give a lag of UVW1 by \us\, of $\sim$150s, i.e. \us\, lagging the X-rays by 580s. However there is sometimes an additional lag between the X-rays and far-UV which would reduce the predicted UVW1 to \us\, lag. On the other hand, the contribution from the BLR may increase the lag slightly. Thus we take a round figure of 600s for the X-ray to \us-band lag. Although this figure is almost certainly not absolutely correct, it probably is not very wrong. We discuss the implications of incorrect referencing in  Section~\ref{sec:models}.  Correct referencing of the HiPERCAM lags relative to the X-ray band is important and is the subject of further observations (Beard et al, in preparation).

\begin{table*}
\begin{tabular}{llllll}
Band &Central &JAVELIN Lag & JAVELIN Lag & Centroid ICCF & Centroid ICCF \\
     &Wavelength &Uncorrected & Corrected   & Uncorrected   & Corrected \\
     &(Angstroms) &           &             &               &           \\ 
{\it $u_{s}$}& 3694  &  0 & 0 & 0 & 0 \\ 
{\it $g_{s}$}& 4778 & $325^{+9}_{-7}$ &  $360^{+10}_{-15}$ & $377 \pm 20$  &  $390 \pm 18$ \\
{\it $r_{s}$}& 6201 & $578^{+12}_{-5}$ & $578^{+11}_{-5}$  & $565 \pm 20$  &  $585 \pm 19$\\ 
{\it $i_{s}$}& 7640 & $606^{+13}_{-3}$ & $633^{+15}_{-14}$ & $571 \pm 23$  &  $615 \pm 23$\\ 
{\it $z_{s}$}& 9066  & $676^{+10}_{-16}$ & $693^{+10}_{-8}$ & $650 \pm 30$  &  $698 \pm 33$    
\end{tabular}
\caption{Lags, relative to the \us\, band, in seconds for the first 8000s of the HiPERCAM observations. The first column gives the band. The second column gives the 
effective, or Bandpass Average wavelength of the band.
The third and fourth columns are the lags as measured by JAVELIN using the original lightcurves and using the seeing-corrected lightcurves respectively. The fifth and sixth columns are the centroid interpolation cross correlation function 
(ICCF) lags as measured using the FR/RSS method \citep{peterson98} on the original lightcurves and on the seeing-corrected lightcurves respectively.
}
\label{tab:lags}
\end{table*}
 
\section{Discussion}
\label{sec:models}

\subsection{Preliminary Conclusions: An Edge to the Emission Region and little BLR contribution}

\label{sec:preliminary}

Our best estimates of the multiwavelength lags in NGC~4395 are given in Fig.~\ref{fig:bigM_rout}.
Even without any comparison with models, two important conclusions can be drawn immediately. Firstly the \gs-band lags considerably behind the \us-band, unlike in almost all other AGN. This observation implies that the BLR is not a major contributor to the UV/optical variability of NGC4395. Secondly, the lag spectrum flattens at long wavelengths, implying an edge to the emission region. These two observations will be considered in more detail below.

\subsection{Full Response Modelling of the Lags: The Disc Edge and The Colour Correction Factor}

\cite{kammoun21_model} give a formula from which it is possible to estimate lags, for a given mass, accretion rate and spin. However that formula is not valid for masses below $10^{6}$\msun so we do not use it here. Instead we used the full KYNxilrev code\footnote{https://projects.asu.cas.cz/stronggravity/kynreverb}, previously used by \cite{kammoun21_model},
to produce response functions at a number of different wavelengths over our observed lag wavelength range. We make the standard assumption that the lag is defined as the mean of the response function. We do have an in-house code (Veresvarska, see Appendix) which provides broadly similar results but, unlike the KYNxilrev code, it does not contain General Relativistic modifications. Hence, as it is also not yet publicly available, we base our results on the KYNxilrev code. 

Besides the mass, other parameters which are required to model lags are the illuminating X-ray luminosity and the accretion rate.

Regarding the illuminating luminosity, although there is considerable variability on short timescales, observers \citep[e.g.][and Vincentelli et al, in prep] {vaughan05_4395, kammoun19_4395} agree on an average 2-10 keV XMM PN count rate of $\sim 0.5$ c/s. \cite{kammoun19_4395} give a 3-10 keV luminosity of $0.9-2.4 \times 10^{40}$ ergs/s and a photon index of 1.6. Taking a mean value, extrapolation to 14-195 keV agrees closely with the luminosity from the BAT 105 month survey\footnote{\url{https://swift.gsfc.nasa.gov/results/bs105mon/}} of $8 \times 10^{40}$ ergs/s. Extrapolating the BAT luminosity to the whole illuminating X-ray band gives $1.6 \times 10^{41}$ ergs/s which is the value we take here.

The accretion rate through the disc, which affects the structure and the lags, is harder to define. There are a number of estimates of the bolometric luminosity, derived from optical measurements via far-from-certain bolometric correction factors, which broadly agree. For example \cite{peterson05} gives $5.4 \times 10^{40}$ ergs/s, \cite{moran99} gives $5.2-8.8 \times 10^{40}$ ergs/s and \cite{brum19_4395} gives $9.9 \times 10^{40}$ ergs/s. However NGC4395 has a larger X-ray/optical ratio than for the higher mass AGN which have been monitored recently with Swift and may well have a jet contribution to the X-ray luminosity \citep[e.g.][]{king13}, so the bolometric corrections may not be accurate. For example the BAT X-ray/g-band ratio of NGC4395 is $3\times$ larger than for the higher mass and higher accretion rate Seyfert 1 galaxy NGC4593 \cite{mch18}. \cite{kammoun19_4395} include high energy (NuSTAR) observations together with optical and derive a bolometric luminosity for NGC~4395 of $1.9\times 10^{41}$ ergs/s.  

If we derive the Eddington accretion ratio from the ratio of bolometric to Eddington luminosity, which bolometric luminosity should we use? As it is possible that material could accrete onto the black hole through some corona over the disc without releasing gravitational energy through the disc, it could be argued that a bolometric luminosity derived from optical observations which are dominated by disc emission, is the most appropriate. This is not a question for this paper but here we take the commonly used value from \cite{peterson05}, corresponding to an Eddington ratio of 0.12 per cent. Choosing a larger value would increase the model lags, however choosing a higher spin would decrease the lags so there are many uncertainties. 

The flattening of the lag spectrum at long wavelengths requires an edge to the emission region which we model here as a truncation in the outer disc radius; see also Fig.22 of \cite{kammoun21_model}. However the shape and amplitude of the lag spectrum also depends greatly on the assumed disc colour correction factor. This factor takes account of the fact that at high temperatures, in the inner disc, hydrogen and helium will be ionised so that the deeper, and hotter, parts of the disc are visible. So the disc will appear hotter than expected on the basis of the standard \cite{shakura73} and \cite{novikov73} models. The colour correction factor is discussed at length by \cite{done12} who provide prescriptions for the correction factor in different temperature regimes. We refer to their standard prescription as the {\sc DONE} colour correction factor. This factor is approximated as 2.4 for $T>10^{5}$K, 1 for $T<10^{4}$K and decreases between those two temperatures.

Finding the most appropriate combination of mass, accretion rate, spin, truncation radius, colour correction factor and even inclination is not trivial or unambiguous. 
Thus our aim here is not to find precise best-fit values but to find a range of acceptable values.
We begin by testing the values of mass ($3.6 \pm 1.1 \times 10^{5}$\msun\, with 1$\sigma$ uncertainty) and accretion rate (0.12 percent of Eddington) given by \cite{peterson05}. This mass is the same as that ($4^{+8}_{-3} \times 10^{5}$\msun\, with quoted 3$\sigma$ uncertainties) derived by \cite{denbrock15} from detailed gas dynamical modelling. We take an intermediate inclination of 45 degrees, which is in good agreement with the value of 37 degrees estimated by \cite{denbrock15} and gives almost identical lags. We assume a lampost X-ray emission geometry with a source height of 10\rg. As noted earlier, changing this value even by a factor of 2 makes little difference to the predicted lags. We also choose zero spin.
Thus the remaining variables are truncation radius and colour correction factor.

\begin{figure}
\hspace{-10mm}
\includegraphics[width=70mm,height=85mm,angle=270]{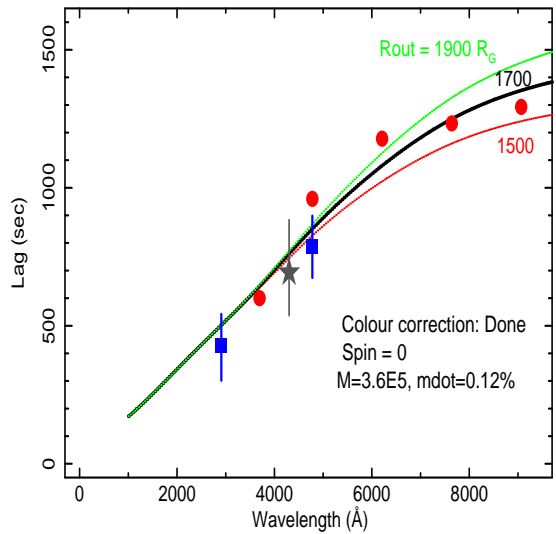}
\caption{The lag observations are as described in Fig.~\ref{fig:lags}. Here we assume a mass of $3.6 \times 10^{5}$\msun, \me=0.12 per cent Eddington, X-ray source height 10\rg, zero spin and the {\sc DONE} colour correction factor.  The lines are lag predictions for different disc outer radii of 1500, 1700 and 1900\rg.
}
\label{fig:bigM_rout}
\end{figure}

Following some experimentation, it became clear that a relatively small truncation radius, $\ltsim 2000$\rg, is required to produce enough flattening of the lags at long wavelengths, almost independent of the other variables. Assuming the {\sc DONE} colour correction factor, which is the best physically motivated factor we show, in Fig.~\ref{fig:bigM_rout}, models for a range of disc outer radii which broadly bracket the observed lag data and have roughly the right shape. A truncation radius of 1600-1700 \rg\, agrees reasonably with the data. 

\begin{figure}
\hspace{-10mm}
\includegraphics[width=70mm,height=85mm,angle=270]{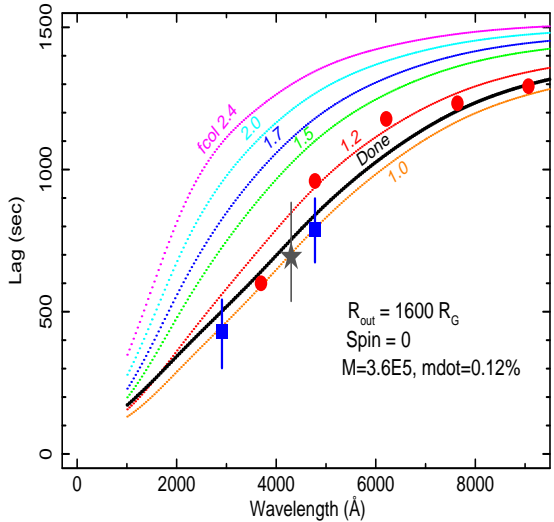}
\caption{As for Fig.~\ref{fig:bigM_rout} but with the disc outer radius fixed at 1600\rg\, and showing a range of colour correction factors, as described in the text.
}
\label{fig:bigM_col}
\end{figure}

To investigate the effect of the colour correction factor, we then fix the outer radius at 1600\rg\, and vary the colour correction factor (Fig.~\ref{fig:bigM_col}) with the other parameters the same as for Fig.~\ref{fig:bigM_rout}. Although the {\sc DONE} prescription varies the colour correction factor as a function of disc temperature, the alternative implementations simply apply the same colour correction factor over the whole disc. A factor of 2.4 is used, following \cite{ross92}, by \cite{kammoun21_model}. A broadly similar correction is derived, though with a different physical model, by \cite{petrucci18}.

The model using the {\sc DONE} colour correction factor, which is very close to almost no colour correction factor (i.e. fcol=1 or 1.2), is closest to the data for this mass, accretion rate and spin. The lag models with higher values of fcol differ greatly from the data, having quite different shapes. \cite{davis19_fcol} give a prescription for fcol as a function of mass, accretion rate and stress parameter. For both the large and small masses, and for values of alpha between 0.001 and 1, fcol remains close to unity, ie close to the {\sc DONE} value.

We have not shown model fits for maximum spin as the lags for maximum spin are $\sim20$ per cent less than those for zero spin and so, for this mass and accretion rate, lie well below the observed lags. 

To perform a brief self-consistency check on our preferred model we have used the {\sc kynsed} model within {\sc XSPEC} to compare the average fluxes of the variable part of the HiPERCAM lightcurves to a predicted SED. It is not our intention here to perform a detailed fit to the SED but simply to perform a consistency check. We will return to this topic in more detail in a future paper.
All input parameters, including distance, are as defined above but the colour correction factor and disc outer radius were left free.  Despite the uncertainties in the fixed parameters such as spin, accretion rate and mass and uncertainties in deriving the input fluxes, the agreement is remarkably close with a p value of 0.10. The derived colour correction factor (assumed to apply over the whole disc) and outer disc radius were $1.4 \pm 0.35$ and $1850 \pm 1900$ \rg\, respectively. Although the disc outer radius result is not terribly constraining, these results are at least consistent with the values derived from the lags.

\begin{figure}
\hspace{-10mm}
\includegraphics[width=70mm,height=85mm,angle=270]{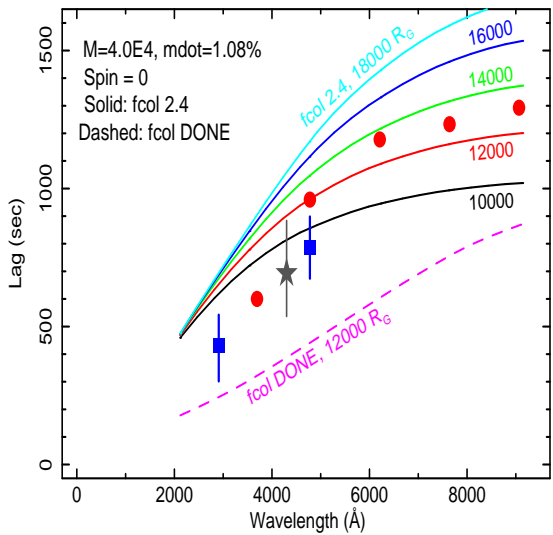}
\caption{The lag observations are as for Fig.~\ref{fig:lags}. The model lines are for a mass of $4.0 \times 10^{4}$\msun, \me=1.08 per cent Eddington, zero spin. The solid lines are for a colour correction factor of 2.4 and the dashed line is for the {\sc DONE} colour correction factor. A range of truncation radii are noted.
}
\label{fig:smallM_rout}
\end{figure}

As a smaller mass, $\sim 4 \times 10^{4}$\msun has been proposed \citep{edri12_4395} we also show, in Fig.~\ref{fig:smallM_rout}, some model lags for that mass with appropriate accretion rate of 1.08 per cent of Eddington. Even for zero spin and a very large truncation radius, the observed lags are almost twice those for the {\sc DONE} correction factor. We therefore experimented with the large colour correction factor, 2.4, used by \cite{kammoun21_model}. We show a range of cutoff radii. 
Although not lying as close to the data as the low colour correction factor models for the larger mass, the colour correction 2.4 model with cutoff radius of 14000\rg, is in approximate agreement with the data. 

We conclude from this preliminary investigation that disc reprocessing models based on the most commonly accepted mass of $3.6 \times 10^{5}$\msun\,  for zero spin and for the most physically motivated disc colour correction factor ({\sc DONE} factor),
are in reasonable agreement with the observed lags but the disc must be truncated with 
an outer disc truncation radius of $\sim1700$\rg.

Models based on a smaller mass, $4 \times 10^{4}$\msun, can also be brought into moderate agreement with the observed lags but do require a very high colour correction factor of 2.4 over the whole disc, not just the inner hot part, and an outer truncation radius of $\sim14000$\rg.

\subsection{BLR contribution to the UV/optical variability}

In almost all other well studied AGN, with the possible exception of Mrk817 \citep{kara21}, the \us-band lag is longer than, or at the very least about the same as, that in the surrounding wavebands. This \us-band excess is usually attributed to a lag contribution from reprocessing in the more distant BLR \citep{korista01,korista19}. 
Here, however, the \us-band lag is considerably shorter than that in the \gs-band. This result depends only on HiPERCAM observations so is completely robust. Although a weak contribution from the BLR is still consistent with the data, 
and the lag from the BLR can be reduced by a suitable choice of gas density and ionisation parameter \citep{lawther18},
the conclusion here is that the disc rather than the
BLR is the dominant contributor to the UV/optical variability, at least on the few hour timescales observed here, in NGC~4395.

An HST UV spectrum of NGC~4395 is presented by \cite{filippenko93_4395} and optical spectra by \cite{lira99_4395}. Most of the emission lines are dominated by an unresolved core but some of the permitted lines have weak broad bases. Line fluxes are given in Tables 2 and 3 from \cite{lira99_4395}. For example, the flux from the broad component of $H_{\beta}$ was observed between 14 and 18$\times 10^{-15}$ \ecs, compared with between 22 and 28$\times 10^{-15}$ \ecs\, for the narrow component, both of which variations are within the quoted calibration uncertainties of 30 per cent. Typical continuum levels below 4000 \AA\, have varied between 0.5 and 1$\times 10^{-15}$ \ecsa. Thus when considered over a typical {\it ugriz} bandpass of $\sim1000$\AA, emission line variability is not a large contributor to the total flux variations. This limited contribution from the BLR may explain why we do not see the large excess $u$-band lag in NGC~4395.

\subsection{Physical Origin of the Truncation Radius}

There are two natural radii which might be considered: the self-gravity radius and the dust sublimation radius. Considering first the self-gravity radius, the physics here is not straightforward and simple. As one moves out in the disc the first change is when the gravity of disc material to itself exceeds the forces supporting the disc, e.g. gas and radiation pressure. Disc structure may become more fragmented but there will still be a disc, available for reprocessing. Still further out cooling may become efficient enough that the disc will collapse into small clumps which will not be effective reprocessors. 

There is considerable modelling of this behaviour which depends on a number of assumptions, e.g. regarding metallicity, illumination and viscosity, which we shall not attempt to repeat. We refer the interested reader to \cite{gammie01} and also to \cite{shore82,laor89,collin-souffrin90,collin-hure99,lobban22,derdzinski22}. Those papers provide formulae enabling the reader to estimate the self-gravity radius. However, of relevance to NGC4395 is that \cite{shore82} note that the 'cliff' which appears at the self-gravity radius, beyond which the disc scale height is much reduced and reprocessing from the disc will stop does not appear for masses $<3 \times 10^{7}$\msun. For lower masses, only a smoothly flared disc results.  We conclude that, although further work is needed to be certain, the self-gravity radius may not be so important in defining an effective outer radius to the disc for reprocessing in low mass AGN like NGC~4395 and we do not consider it further.

Considering next the sublimation radius, \cite{baskin18} state that, at the radius at which the disc temperature falls to below the dust sublimation temperature, the atmosphere of the disc will become dusty, its opacity will greatly increase and the disc can again become radiation-pressure supported. The resultant inflated dusty disc obscures the region behind it, leading to an edge to the part of the disc which can reprocess radiation. 
A dusty disc wind, driven by radiation pressure, will also form, which would constitute the inner radius of the BLR and give rise to the wind component seen in some emission lines, such as CIV \citep[e.g.][]{coatman16}.
The dust sublimation radius is therefore a natural edge to reprocessing either from the disc or the BLR. 

The dust sublimation radius, \rsub, depends on the dust composition and on the temperature at which it will sublime. Taking a sublimation temperature of 2000K 
\cite{baskin18} with an X-ray source height of 10 \rg\ an illumination luminosity of $1.6 \times 10^{41}$ \ecs, an albedo of 0.2 and zero spin we calculate sublimation radii for dust on the surface of the illuminated disc of
4300\rg\, and 19000\rg\, respectively for the large and small masses. 
For the smaller mass this radius is not too much larger than the radius determined from the lags, and may be consistent within the many uncertainties of the parameters, but for the large mass the sublimation radius is larger. For the small mass a dusty wind might therefore obscure the outer part of the disc and provide a suitable edge.

For the larger mass the sublimation radius is a factor of almost 3 further out than the truncation radius derived from the lags. We might try to drop the sublimation radius by modelling the lags with a maximum spin disc. However for maximum spin and the same observed Bolometric luminosity, i.e. the same ratio of $L_{Bolometric}/L_{Eddington}$, which is commonly referred to as the Eddington accretion rate, the required accretion rate in kg/sec then drops. This drop occurs as much of the luminosity, unlike for zero spin, is then emitted within 6\rg. Thus given a lower accretion rate in kg/s, the outer disc becomes cooler and, in the present case, the sublimation radius drops by 16 per cent. However, given the addition of reprocessing material within 6\rg, the lags also drop (here by about 10 per cent), so that does not help much.
The sublimation radii of cooler, unilluminated discs are lower, closer to the truncation radii measured from the lags, but we require illumination to produce the variable UV/optical emission. Also, for an almost unilluminated disc, the model lags drop by almost a factor of 2, at least at short wavelengths. If the disc continues out to the sublimation radius and maybe beyond,  we therefore require material to shield the outer portion of the disc from the central illumination. If this material is in the form of a disc wind we would need it, for the large mass, to leave the disc at a local temperature of $\sim 4-5000$K, i.e. above the dust sublimation temperature. Probably this material would have to be gaseous rather than dusty as, once raised from the disc and exposed fully to the illuminating radiation, the dust would be rapidly vapourised.  A line-driven wind might be applicable in this case \citep[e.g.][]{proga04_winds,higginbottom14}.
We comment further in Section~\ref{sec:other}.

\subsection{Alternative explanations of the shorter than expected long wavelength lags}
\label{sec:other}

The \is\, and \zs\, lags here are shorter than expected from disc reprocessing. 
Shorter than expected lags do not necessarily imply reprocessing material closer to the illuminating source than expected, we simply require that the path from the illuminating source, via the reprocessor to the observer is shorter than expected. A wind, close to the line of sight \cite[e.g.][]{honig19} might, depending on the orientation, add very little extra distance for the light path from the central source to the observer and might produce very short near-IR lags. The base of the wind, if optically thick, might provide an effective wall which might mimic a truncated disc, at least for the longer wavelengths.

The model AGN lag spectra from the BLR by \cite{netzer22} contain a peak due to the Paschen continuum at 8200\AA\, and then become shorter at longer wavelengths. Although this decrease, compared to the expectation from a standard untruncated disc, is in the same sense as our observed decrease, the change which we observe is even more extreme. However a difficulty with interpreting the present observed lag spectra as being mainly due to reprocessing in the BLR is that here the \gs-band lags a long way after the \us-band, whereas for BLR reprocessing we expect those two lags to be the other way around.

However using the relationship between the $H_\beta$ BLR radius and the 5100\AA\, luminosity for low luminosity AGN derived by \cite{bentz13}, and taking a luminosity of $5.9 \times 10^{39}$ ergs s$^{-1}$ from \cite{peterson05}, we derive a radius, in light-travel time, of 1600s. Given the many uncertainties involved, this figure is quite close to the lag at long wavelengths which our measured lags seem to be asymptotically approaching. Thus the observations are consistent with the inner edge of the BLR being at approximately the same radius as the outer edge of the emission region, implying a link between those two structures.

A possible scenario then is that an  optically thick wind arises from the disc which both shields the outer regions of the disc from central illumination and acts, itself, as a reprocessor. If the disc temperature is above the dust sublimation temperature this may not be a dusty wind but winds can be driven by other mechanisms, eg line driving, leaving the disc at higher temperatures than the sublimation temperature. 
The base of such winds can be optically thick and can shield the outer disc from X-ray irradiation \citep{dehghanian20_agnstormxi,higginbottom14,proga04_winds}.
The wind may also act as a reprocessor and contribute to the variable UV/optical emission and there may even be some feedback of radiation from the wind onto the disc. 
Depending on the height of the centroid of the X-ray emission above the black hole, and the flaring profile of the disc, the optically thick part of the disc need not be particularly high. Also, as noted by \citep{dehghanian20_agnstormxi}, the wind base might not be optically thick all the time, hence lag spectra may change with time. At greater heights, the optically thinner wind might become the BLR. However the broad components to the optical emission lines in NGC4395 are very weak and so if the wind becomes the BLR is must becomes optically very thin at quite low heights, providing a small effective solid angle for reprocessing of line emission. None of these effects are considered here and clearly, further modelling is needed to test the wind scenario. 

\subsection{Implications for other AGN}

The observations, and possible interpretation, presented here have implications for our understanding of the previously puzzling lag spectra of some other AGN.

\subsubsection{The X-ray / UV disconnect}

As noted in the Introduction, one of the major problems in interpreting AGN lag spectra has been the observed excess lag between the X-rays and the shortest wavelength UV band, compared to lags between the UV and optical bands. 
Whilst it may not be the only cause of this excess lag, fitting a disc model without truncation to observations from a disc that actually is truncated will automatically produce an apparent excess lag.  A truncated disc model is then consistent with a direct view of the disc from the central X-ray source rather than requiring additional scattering through, e.g., an inflated inner edge of the disc \citep{gardner17}, to explain the excess X-ray to UV lag.

\subsubsection{Overall lag modelling and mass estimation}

As we can see from Figs.~\ref{fig:bigM_rout} and ~\ref{fig:smallM_rout}, if the UV/optical variability really does originate from reprocessing of X-ray emission by a truncated accretion disc, fitting a model from an un-truncated disc will produce nonsense and completely incorrect mass estimates. The very short wavelength bands, being the least affected by disc truncation, are the most sensitive to the mass and accretion rate. 
Here we note that response functions for longer wavelength bands than B show a cliff
at long timescales but the shorter wavelength responses are unaffected.
The red bands are most sensitive to the truncation radius. 
A flattening of the lag spectrum with increasing wavelength is an indication of disc truncation so it is important to define the lag spectral slope well at short wavelengths.
\cite{cackett18}, using HST to extend lag measurements down to $\sim$1100\AA, finds a steepening at wavelengths below 2000\AA\, in NGC4593 which is consistent with disc truncation.
With only a small number of bands, there will be degeneracy in the fits but with a full range of wavelengths quite sensitive mass estimates, and disc truncation radii, may be made.

If there is only a mild degree of truncation, affecting only the longest wavelengths, then lag fits of the form  $lag \propto \lambda^{\beta}$, which do not include any truncation, could produce values of  $\beta$ nearer to unity than 4/3, e.g. as seen in NGC5548 \cite{fausnaugh16}.

\section{Conclusions}

We present three nights ($\sim 15$hr) of {\it u, g, r, z} and two nights ($\sim 12$hr) of {\it u, B, g} imaging of the low mass AGN NGC~4395 on the robotic LT. These data provide lightcurves which show clear correlated variations on few hour timescales but with interband lags too short to be measured accurately given the 
$\sim200$s sampling. A follow-up 6~hr observation with HiPERCAM on the 10.4m GTC provided superb simultaneous 5-band (\us, \gs, \rs, \is\, and \zs) imaging, with 15s sampling in \us~ and 3s in all other bands, the fastest yet reported for an AGN. The observation is dominated by two smooth, almost sinusoidal, variations, similar to those seen in the LT observations.

Very clear correlated variability is seen in all HiPERCAM bands and wavelength dependent lags between bands are measured relative to the \us-band with much higher precision than in any other AGN. Two main points are immediately obvious from the lags measured from the HiPERCAM observations. Firstly, and
unlike in almost all other AGN, the \gs-band lags considerably after the \us-band. Thus the \us-band is not dominated by Balmer continuum emission from a distant BLR and hence reprocessing in the BLR is not a large contributor to the UV/optical continuum variability in this AGN. Secondly, the lags between the \rs, \is~ and \zs-bands are very small, so that the lag spectrum flattens off at long wavelengths. These observations imply an edge to the emission region which limits the outward movement of the centroid of the emission regions at long wavelengths. A likely possibility is a truncated accretion disc. A brief investigation shows that the observed SED based on the variable component in the HiPERCAM observations is consistent with this model.

We observe that when the source is brighter it is bluer and also the lags, at least at short wavelengths, are longer. These results are in fairly good agreement with the expectation of disc reprocessing with a higher illuminating luminosity, around a factor 2 or 3. Variations in the X-ray luminosity by even larger factors on comparable timescales are commonly seen in X-ray observations \citep{vaughan05_4395} and could easily explain such variations. We hope to repeat and test this analysis with higher quality HiPERCAM data.

We have combined the HiPERCAM lags with lags measured previously between the X-ray, UV and $g$ band \citep{mch16} and have fitted the combined lag spectra with truncated accretion disc models using lags derived from the KYNxilrev
code and also from our own in-house models. Similar results are obtained with all codes. We have not tried to perform a full statistical fit over all parameters but have compared the observed lag spectrum with models based on the most commonly assumed mass of $3.6 \times 10^{5}$\msun\, \citep{peterson05,denbrock15} and also on a mass a factor of 10 lower \citep{edri12_4395}. 

We demonstrated that the disc colour correction factor has a large effect on the model lags. The {\sc DONE} temperature-dependent factor is probably the best physically motivated correction factor and is also similar to the colour correction factor near unity which can be derived for NGC~4395 using the formulation from \cite{davis19_fcol}. Using that factor
we find that model lags based on a truncation radius of $\sim1700$\rg\, agree with the observed lags reasonably well for the mass of $3.6 \times 10^{5}$\msun and zero spin. For a mass of $4 \times 10^{4}$\msun\, the model lags based on the {\sc DONE} colour correction factor are too short. Although they are not quite as close as for the {\sc DONE} factor with the larger mass, approximate model agreement with the observed lags can be obtained for the smaller mass but only if a colour correction factor of 2.4 is applied over the whole disc, which is physically implausible.

Regarding possible reasons for the truncated disc, \cite{shore82} note that for low masses such as those considered here the disc does not suffer catastrophic collapse at large radii due to self-gravity effects and so self-gravity is probably not the cause of the truncated disc. For the small mass the disc truncation radius measured from the observed lags is about at the same radius as the dust sublimation radius. However for the large mass the sublimation radius is almost a factor 2-3 larger than the truncation radius measured from the lags. If our calculations are correct, a shield of some sort is required to cut off illumination to the outer disc. A dusty disc wind might work for the smaller mass but for the larger mass, where the dust sublimation radius lies beyond the measured disc truncation radius, a different type of wind, eg line-driven, might be more applicable \citep{proga04_winds,higginbottom14}.

We note that, if the relationship between optical luminosity and BLR radius for low luminosity AGN \citep{bentz13} is applicable to NGC4395, the inner edge of the BLR should be at a light radius of 1600s, which is close to the maximum lag measured from our data. Thus the inner edge of the BLR may be linked to a wind which is
at least partially responsible for the apparently truncated disc.

There are a number of caveats. In particular there are a large number of variables involved in the lag modelling and we have not carried out an extensive parameter search. It may very well be that a different combination of variables would produce model lags which are closer to the observed lags.  We note in particular the uncertainty regarding the effective accretion rate. If a higher value, derived using the full bolometric luminosity, is applied, the model lags for the high mass option lie well above the observed lags and the small mass option becomes a more attractive alternative. However we do not believe that more detailed lag modelling is yet merited by the data.

We also note that although the lags between the HiPERCAM bands are determined with very high precision, we only have one set of measurements and the observations should be repeated to test for long term stability. We also caution that referencing the HiPERCAM lags to the X-ray band requires an estimate of the lag between the XMM-Newton UVW1 band (2910\AA) and the HiPERCAM \us-band (3695\AA). Due to the closeness of these wavebands the lag is unlikely to be large ($\ltsim150$s), but it is not zero. We also only have one measurement of a lag between the X-rays and any UV band, here UVW1. However any change in the registration between the X-ray frame and the HiPERCAM frame would mainly result only in slightly changing the best estimate of the truncation radius and cannot affect the main conclusions of this paper. To address the registration concern,
further X-ray and UV observations from four full XMM-Newton orbits ($>400$ks), together with ground based $g$-band observations, have been made and will be reported soon (Beard et al, paper 1 in prep).

We note that, immediately prior to submission, we have been informed that another group, using different methods, are working on a similar conclusion regarding the outer edge of the disc (Starkey et al, private communication).

We finally note that the HiPERCAM observations define the optical high frequency power spectrum very well. A paper describing the overall optical power spectrum, including long-timescale (months-year) monitoring from the LT and intermediate scale (hours-weeks) observations from TESS \citep{burke20_4395} is also in preparation (Beard et al, paper 2). 

\section*{Acknowledgements}

We dedicate this paper to the memory of our friend and colleague Tom Marsh who very sadly died whilst this paper was being refereed. As well as making excellent contributions in many other areas of astronomy, Tom was a critical member of the team that developed the HiPERCAM instrument, on which this paper is based. He will be greatly missed. 

We thank the referee, Patrica Ar\'{e}valo, for an extremely helpful and constructive response which greatly improved the quality of the paper.

This paper is based on observations made with the Gran Telescopio Canarias (GTC) and with the Liverpool Telescope (LT). Both the GTC and the LT are installed at the Spanish Observatorio del Roque de los Muchachos of the Instituto de Astrof\'\i sica de Canarias, on the island of La Palma. The LT is operated by Liverpool John Moores University with financial support from the UK Science and Technology Facilities Council.

We thank Michal Dovciak for assistance in getting the KYNxilrev
software to run properly. We also thank Iossif Papadakis, Ari Laor and Hagai Netzer for useful discussions over a number of disc, BLR and black hole mass issues.

The design and construction of HiPERCAM was funded by the European Research Council under the European Union's Seventh Framework Programme (FP/2007-2013) under ERC-2013-ADG Grant Agreement No. 340040 (HiPERCAM). VSD and HiPERCAM operations are supported by STFC grant ST/V000853/1. 

IMcH and FMV thank STFC for support under grant ST/J001600/1 and EB thanks STFC for support under grant ST/S000623/1.  TRM thanked STFC for support under grant ST/T000406/1.
MB and MV thank STFC for studentships under numbers ST/S505705/1 and ST/W507428/1 respectively.

EK acknowledges financial support from the Centre National d’Etudes Spatiales (CNES). 
JHK. acknowledges financial support from the State Research Agency
(AEI-MCINN) of the Spanish Ministry of Science and Innovation under the
grant "The structure and evolution of galaxies and their central
regions" with reference PID2019-105602GB-I00/10.13039/501100011033, from
the ACIISI, Consejer\'{i}a de Econom\'{i}a, Conocimiento y Empleo del
Gobierno de Canarias and the European Regional Development Fund (ERDF)
under grant with reference PROID2021010044, and from IAC project
P/300724, financed by the Ministry of Science and Innovation, through
the State Budget and by the Canary Islands Department of Economy,
Knowledge and Employment, through the Regional Budget of the Autonomous
Community. 

\section*{Data Availability}
The data used in this paper are available from the public archives of the 
LT\footnote{\url{https://telescope.livjm.ac.uk/cgi-bin/lt_search}}, proposal PL17A21 and GTC\footnote{\url{https://gtc.sdc.cab.inta-csic.es/gtc/jsp/searchform.jsp}}, proposal GTC46/18A. Lightcurves can be produced from these data using the HiPERCAM software\footnote{\url{http://deneb.astro.warwick.ac.uk/phsaap/hipercam/docs/html/}}.
For the purpose of open access, the authors have applied a creative commons attribution (CC BY) licence to any author-accepted manuscript version arising.

\section*{Appendix}

\subsection*{Comparison of KYNxilrev lags with our in-house lags}
\begin{figure}
\hspace{-10mm}
\includegraphics[width=70mm,height=85mm,angle=270]{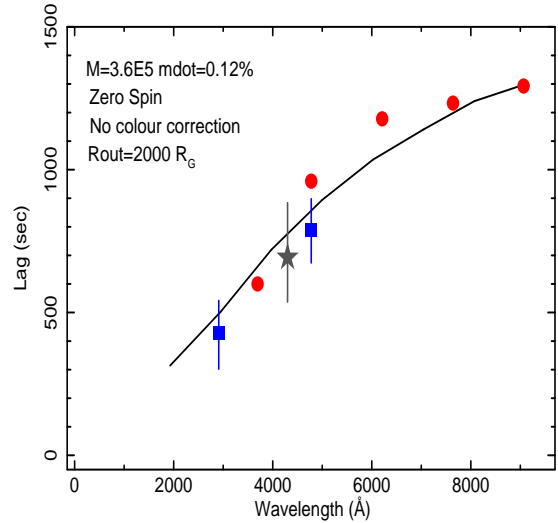}
\caption{As for Fig.~\ref{fig:bigM_rout} but the model lags are from our present in-house code.}
\label{fig:martina_rout}
\end{figure}

In Fig.~\ref{fig:martina_rout} we present, for comparison, lags from the most recent version of our in-house code (Veresvarska, in prep). This version uses the \cite{novikov73} disc temperature prescription but does not include any General Relativistic ray tracing to compute path lengths in the very inner disc. The parameters used are almost identical to that used in 
Figs.~\ref{fig:bigM_rout} and \ref{fig:bigM_col} except that here we apply no colour correction (although the code does allow colour correction) and, by eye, a slightly larger truncation radius of 2000 \rg\, rather than 1700 is closest to the data.  An albedo of 0.2 is used here whereas for Fig.~\ref{fig:bigM_rout} the fraction of illuminating radiation absorbed is calculated using detailed X-ray reflection modelling within the KYNxilrev code. The two model lags are very similar.

\begin{figure}
\hspace{-10mm}
\includegraphics[width=70mm,height=85mm,angle=270]{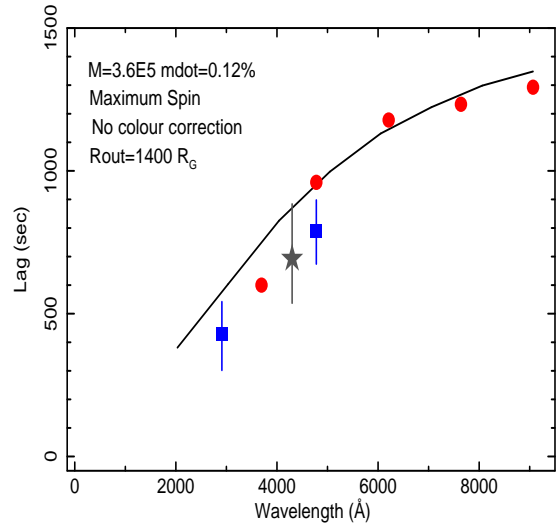}
\caption{Data as for Fig.~\ref{fig:bigM_rout} but the model lags are from our original in-house code.}
\label{fig:sam}
\end{figure}

We also present, in Fig.~\ref{fig:sam}, model lags from the code which used in previous publications \citep[e.g.][]{mch18}. This code uses the \cite{shakura73} disc temperature
prescription. Here a slightly smaller truncation radius, 1400 \rg\, and maximum spin provides a closer 'by eye' fit. The lags, in this model, for zero spin are 20 percent larger, thus are not enormously different to those shown. Again the model lags are close to those of KYNxilrev.

The aim here is not to carry out detailed comparison between the KYNxilrev code and our original in-house code but simply to show that the main result presented here, i.e. that modelling the lags as resulting from disc reprocessing requires a truncated disc, does not depend significantly on any particular modelling code. 

\begin{thebibliography}{82}
\expandafter\ifx\csname natexlab\endcsname\relax\def\natexlab#1{#1}\fi

\bibitem[{{Ar{\'e}valo} {et~al}\mbox{.}(2008){Ar{\'e}valo}, {Uttley}, {Kaspi},
  {Breedt}, {Lira}, \& {M$\rm^{c}$Hardy}}]{arevalo08_2251}
{Ar{\'e}valo} P., {Uttley} P., {Kaspi} S., {Breedt} E., {Lira} P.,
  {M$\rm^{c}$Hardy} I.~M., 2008, \mnras, 389, 1479

\bibitem[{{Ar{\'e}valo} {et~al}\mbox{.}(2009){Ar{\'e}valo}, {Uttley}, {Lira},
  {Breedt}, {M$\rm^{c}$Hardy}, \& {Churazov}}]{arevalo09}
{Ar{\'e}valo} P., {Uttley} P., {Lira} P., {Breedt} E., {M$\rm^{c}$Hardy} I.~M.,
  {Churazov} E., 2009, \mnras, 397, 2004

\bibitem[{{Baskin} \& {Laor}(2018)}]{baskin18}
{Baskin} A., {Laor} A., 2018, \mnras, 474, 1970

\bibitem[{{Bentz} {et~al}\mbox{.}(2013){Bentz}, {Denney}, {Grier}, {Barth},
  {Peterson}, {Vestergaard}, {Bennert}, {Canalizo}, {De Rosa}, {Filippenko},
  {Gates}, {Greene}, {Li}, {Malkan}, {Pogge}, {Stern}, {Treu}, \&
  {Woo}}]{bentz13}
{Bentz} M.~C. {et~al.}, 2013, \apj, 767, 149

\bibitem[{{Berkley} {et~al}\mbox{.}(2000){Berkley}, {Kazanas}, \&
  {Ozik}}]{berkley00}
{Berkley} A.~J., {Kazanas} D., {Ozik} J., 2000, \apj, 535, 712

\bibitem[{{Blandford} \& {McKee}(1982)}]{blandford_mckee82}
{Blandford} R.~D., {McKee} C.~F., 1982, \apj, 255, 419

\bibitem[{{Breedt}(2010)}]{breedt10_thesis}
{Breedt} E., 2010, PhD thesis, Physics and Astronomy, University of Southampton

\bibitem[{{Breedt} {et~al}\mbox{.}(2009){Breedt}, {Ar{\'e}valo},
  {M$\rm^{c}$Hardy}, {Uttley}, {Sergeev}, {Minezaki}, {Yoshii}, {Gaskell},
  {Cackett}, {Horne}, \& {Koshida}}]{breedt09}
{Breedt} E. {et~al.}, 2009, \mnras, 394, 427

\bibitem[{{Breedt} {et~al}\mbox{.}(2010){Breedt}, {M$\rm^{c}$Hardy},
  {Ar{\'e}valo}, {Uttley}, {Sergeev}, {Minezaki}, {Yoshii}, {Sakata}, {Lira},
  \& {Chesnok}}]{breedt10}
{Breedt} E. {et~al.}, 2010, \mnras, 403, 605

\bibitem[{{Brum} {et~al}\mbox{.}(2019){Brum}, {Diniz}, {Riffel},
  {Rodr{\'\i}guez-Ardila}, {Ho}, {Riffel}, {Mason}, {Martins}, {Petric}, \&
  {S{\'a}nchez-Janssen}}]{brum19_4395}
{Brum} C. {et~al.}, 2019, \mnras, 486, 691

\bibitem[{{Burke} {et~al}\mbox{.}(2020){Burke}, {Shen}, {Chen}, {Scaringi},
  {Faucher-Giguere}, {Liu}, \& {Yang}}]{burke20_4395}
{Burke} C.~J., {Shen} Y., {Chen} Y.-C., {Scaringi} S., {Faucher-Giguere} C.-A.,
  {Liu} X., {Yang} Q., 2020, \apj, 899, 136

\bibitem[{{Cackett} {et~al}\mbox{.}(2018){Cackett}, {Chiang},
  {M$\rm^{c}$Hardy}, {Edelson}, {Goad}, {Horne}, \& {Korista}}]{cackett18}
{Cackett} E.~M., {Chiang} C.-Y., {M$\rm^{c}$Hardy} I., {Edelson} R., {Goad}
  M.~R., {Horne} K., {Korista} K.~T., 2018, \apj, 857, 53

\bibitem[{{Cackett} {et~al}\mbox{.}(2020){Cackett}, {Gelbord}, {Li}, {Horne},
  {Wang}, {Barth}, {Bai}, {Bian}, {Carroll}, {Du}, {Edelson}, {Goad}, {Ho},
  {Hu}, {Khatu}, {Luo}, {Miller}, \& {Yuan}}]{cackett20}
{Cackett} E.~M. {et~al.}, 2020, \apj, 896, 1

\bibitem[{{Cackett} {et~al}\mbox{.}(2007){Cackett}, {Horne}, \&
  {Winkler}}]{cackett07}
{Cackett} E.~M., {Horne} K., {Winkler} H., 2007, \mnras, 380, 669

\bibitem[{{Cackett} {et~al}\mbox{.}(2014){Cackett}, {Zoghbi}, {Reynolds},
  {Fabian}, {Kara}, {Uttley}, \& {Wilkins}}]{cackett14}
{Cackett} E.~M., {Zoghbi} A., {Reynolds} C., {Fabian} A.~C., {Kara} E.,
  {Uttley} P., {Wilkins} D.~R., 2014, \mnras, 438, 2980

\bibitem[{{Cameron}(2014)}]{cameron14_thesis}
{Cameron} D., 2014, PhD thesis, Physics and Astronomy, University of
  Southampton

\bibitem[{{Cameron} {et~al}\mbox{.}(2012){Cameron}, {M$\rm^{c}$Hardy},
  {Dwelly}, {Breedt}, {Uttley}, {Lira}, \& {Arevalo}}]{cameron12}
{Cameron} D.~T., {M$\rm^{c}$Hardy} I., {Dwelly} T., {Breedt} E., {Uttley} P.,
  {Lira} P., {Arevalo} P., 2012, \mnras, 422, 902

\bibitem[{{Chartas} {et~al}\mbox{.}(2012){Chartas}, {Kochanek}, {Dai}, {Moore},
  {Mosquera}, \& {Blackburne}}]{chartas12}
{Chartas} G., {Kochanek} C.~S., {Dai} X., {Moore} D., {Mosquera} A.~M.,
  {Blackburne} J.~A., 2012, \apj, 757, 137

\bibitem[{{Coatman} {et~al}\mbox{.}(2016){Coatman}, {Hewett}, {Banerji}, \&
  {Richards}}]{coatman16}
{Coatman} L., {Hewett} P.~C., {Banerji} M., {Richards} G.~T., 2016, \mnras,
  461, 647

\bibitem[{{Collin} \& {Hur{\'e}}(1999)}]{collin-hure99}
{Collin} S., {Hur{\'e}} J.-M., 1999, \aap, 341, 385

\bibitem[{{Collin-Souffrin} \& {Dumont}(1990)}]{collin-souffrin90}
{Collin-Souffrin} S., {Dumont} A.~M., 1990, \aap, 229, 292

\bibitem[{{Dai} {et~al}\mbox{.}(2010){Dai}, {Kochanek}, {Chartas},
  {Koz{\l}owski}, {Morgan}, {Garmire}, \& {Agol}}]{dai10}
{Dai} X., {Kochanek} C.~S., {Chartas} G., {Koz{\l}owski} S., {Morgan} C.~W.,
  {Garmire} G., {Agol} E., 2010, \apj, 709, 278

\bibitem[{{Davis} \& {El-Abd}(2019)}]{davis19_fcol}
{Davis} S.~W., {El-Abd} S., 2019, \apj, 874, 23

\bibitem[{{Dehghanian} {et~al}\mbox{.}(2020){Dehghanian}, {Ferland}, {Kriss},
  {Peterson}, {Korista}, {Goad}, {Chatzikos}, {Guzm{\'a}n}, {De Rosa},
  {Mehdipour}, {Kaastra}, {Mathur}, {Vestergaard}, {Proga}, {Waters}, {Bentz},
  {Bisogni}, {Brandt}, {Dalla Bont{\`a}}, {Fausnaugh}, {Gelbord}, {Horne},
  {McHardy}, {Pogge}, \& {Starkey}}]{dehghanian20_agnstormxi}
{Dehghanian} M. {et~al.}, 2020, \apj, 898, 141

\bibitem[{{den Brok} {et~al}\mbox{.}(2015){den Brok}, {Seth}, {Barth},
  {Carson}, {Neumayer}, {Cappellari}, {Debattista}, {Ho}, {Hood}, \&
  {McDermid}}]{denbrock15}
{den Brok} M. {et~al.}, 2015, \apj, 809, 101

\bibitem[{{Derdzinski} \& {Mayer}(2022)}]{derdzinski22}
{Derdzinski} A., {Mayer} L., 2022, arXiv e-prints, arXiv:2205.10382

\bibitem[{{Dhillon} {et~al}\mbox{.}(2021){Dhillon}, {Bezawada}, {Black},
  {Dixon}, {Gamble}, {Gao}, {Henry}, {Kerry}, {Littlefair}, {Lunney}, {Marsh},
  {Miller}, {Parsons}, {Ashley}, {Breedt}, {Brown}, {Dyer}, {Green},
  {Pelisoli}, {Sahman}, {Wild}, {Ives}, {Mehrgan}, {Stegmeier}, {Dubbeldam},
  {Morris}, {Osborn}, {Wilson}, {Casares}, {Mu{\~n}oz-Darias}, {Pall{\'e}},
  {Rodr{\'\i}guez-Gil}, {Shahbaz}, {Torres}, {de Ugarte Postigo},
  {Cabrera-Lavers}, {Corradi}, {Dom{\'\i}nguez}, \&
  {Garc{\'\i}a-Alvarez}}]{dhillon21}
{Dhillon} V.~S. {et~al.}, 2021, \mnras, 507, 350

\bibitem[{{Done} {et~al}\mbox{.}(2012){Done}, {Davis}, {Jin}, {Blaes}, \&
  {Ward}}]{done12}
{Done} C., {Davis} S.~W., {Jin} C., {Blaes} O., {Ward} M., 2012, \mnras, 420,
  1848

\bibitem[{{Edelson} {et~al}\mbox{.}(2019){Edelson}, {Gelbord}, {Cackett},
  {Peterson}, {Horne}, {Barth}, {Starkey}, {Bentz}, {Brandt}, {Goad}, {Joner},
  {Korista}, {Netzer}, {Page}, {Uttley}, {Vaughan}, {Breeveld}, {Cenko},
  {Done}, {Evans}, {Fausnaugh}, {Ferland}, {Gonzalez-Buitrago}, {Gropp},
  {Grupe}, {Kaastra}, {Kennea}, {Kriss}, {Mathur}, {Mehdipour}, {Mudd},
  {Nousek}, {Schmidt}, {Vestergaard}, \& {Villforth}}]{edelson19}
{Edelson} R. {et~al.}, 2019, \apj, 870, 123

\bibitem[{{Edelson} {et~al}\mbox{.}(2015){Edelson}, {Gelbord}, {Horne},
  {M$\rm^{c}$Hardy}, {Peterson}, {Ar{\'e}valo}, {Breeveld}, {De Rosa}, {Evans},
  {Goad}, {Kriss}, {Brandt}, {Gehrels}, {Grupe}, {Kennea}, {Kochanek},
  {Nousek}, {Papadakis}, {Siegel}, {Starkey}, {Uttley}, {Vaughan}, {Young},
  {Barth}, {Bentz}, {Brewer}, {Crenshaw}, {Dalla Bont{\`a}}, {De
  Lorenzo-C{\'a}ceres}, {Denney}, {Dietrich}, {Ely}, {Fausnaugh}, {Grier},
  {Hall}, {Kaastra}, {Kelly}, {Korista}, {Lira}, {Mathur}, {Netzer},
  {Pancoast}, {Pei}, {Pogge}, {Schimoia}, {Treu}, {Vestergaard}, {Villforth},
  {Yan}, \& {Zu}}]{edelson15}
{Edelson} R. {et~al.}, 2015, \apj, 806, 129

\bibitem[{{Edelson} \& {Krolik}(1988)}]{edelson88}
{Edelson} R.~A., {Krolik} J.~H., 1988, \apj, 333, 646

\bibitem[{{Edri} {et~al}\mbox{.}(2012){Edri}, {Rafter}, {Chelouche}, {Kaspi},
  \& {Behar}}]{edri12_4395}
{Edri} H., {Rafter} S.~E., {Chelouche} D., {Kaspi} S., {Behar} E., 2012, \apj,
  756, 73

\bibitem[{{Emmanoulopoulos} {et~al}\mbox{.}(2014){Emmanoulopoulos},
  {Papadakis}, {Dov{\v c}iak}, \& {M$\rm^{c}$Hardy}}]{emmanoulopoulos14}
{Emmanoulopoulos} D., {Papadakis} I.~E., {Dov{\v c}iak} M., {M$\rm^{c}$Hardy}
  I.~M., 2014, \mnras, 439, 3931

\bibitem[{{Fausnaugh} {et~al}\mbox{.}(2016){Fausnaugh}, {Denney}, {Barth},
  {Bentz}, {Bottorff}, {Carini}, {Croxall}, {De Rosa}, {Goad}, {Horne},
  {Joner}, {Kaspi}, {Kim}, {Klimanov}, {Kochanek}, {Leonard}, {Netzer},
  {Peterson}, {Schn{\"u}lle}, {Sergeev}, {Vestergaard}, {Zheng}, {Zu},
  {Anderson}, {Ar{\'e}valo}, {Bazhaw}, {Borman}, {Boroson}, {Brandt},
  {Breeveld}, {Brewer}, {Cackett}, {Crenshaw}, {Dalla Bont{\`a}}, {De
  Lorenzo-C{\'a}ceres}, {Dietrich}, {Edelson}, {Efimova}, {Ely}, {Evans},
  {Filippenko}, {Flatland}, {Gehrels}, {Geier}, {Gelbord}, {Gonzalez},
  {Gorjian}, {Grier}, {Grupe}, {Hall}, {Hicks}, {Horenstein}, {Hutchison},
  {Im}, {Jensen}, {Jones}, {Kaastra}, {Kelly}, {Kennea}, {Kim}, {Korista},
  {Kriss}, {Lee}, {Lira}, {MacInnis}, {Manne-Nicholas}, {Mathur},
  {M$\rm^{c}$Hardy}, {Montouri}, {Musso}, {Nazarov}, {Norris}, {Nousek},
  {Okhmat}, {Pancoast}, {Papadakis}, {Parks}, {Pei}, {Pogge}, {Pott}, {Rafter},
  {Rix}, {Saylor}, {Schimoia}, {Siegel}, {Spencer}, {Starkey}, {Sung}, {Teems},
  {Treu}, {Turner}, {Uttley}, {Villforth}, {Weiss}, {Woo}, {Yan}, \&
  {Young}}]{fausnaugh16}
{Fausnaugh} M.~M. {et~al.}, 2016, \apj, 821, 56

\bibitem[{{Filippenko} {et~al}\mbox{.}(1993){Filippenko}, {Ho}, \&
  {Sargent}}]{filippenko93_4395}
{Filippenko} A.~V., {Ho} L.~C., {Sargent} W. L.~W., 1993, \apjl, 410, L75

\bibitem[{{Gammie}(2001)}]{gammie01}
{Gammie} C.~F., 2001, \apj, 553, 174

\bibitem[{{Gardner} \& {Done}(2017)}]{gardner17}
{Gardner} E., {Done} C., 2017, \mnras, 470, 3591

\bibitem[{{Hern{\'a}ndez Santisteban} {et~al}\mbox{.}(2020){Hern{\'a}ndez
  Santisteban}, {Edelson}, {Horne}, {Gelbord}, {Barth}, {Cackett}, {Goad},
  {Netzer}, {Starkey}, {Uttley}, {Brandt}, {Korista}, {Lohfink}, {Onken},
  {Page}, {Siegel}, {Vestergaard}, {Bisogni}, {Breeveld}, {Cenko}, {Dalla
  Bont{\`a}}, {Evans}, {Ferland}, {Gonzalez-Buitrago}, {Grupe}, {Joner},
  {Kriss}, {LaPorte}, {Mathur}, {Marshall}, {Mehdipour}, {Mudd}, {Peterson},
  {Schmidt}, {Vaughan}, \& {Valenti}}]{hernandez20}
{Hern{\'a}ndez Santisteban} J.~V. {et~al.}, 2020, \mnras, 498, 5399

\bibitem[{{Higginbottom} {et~al}\mbox{.}(2014){Higginbottom}, {Proga},
  {Knigge}, {Long}, {Matthews}, \& {Sim}}]{higginbottom14}
{Higginbottom} N., {Proga} D., {Knigge} C., {Long} K.~S., {Matthews} J.~H.,
  {Sim} S.~A., 2014, \apj, 789, 19

\bibitem[{{H{\"o}nig}(2019)}]{honig19}
{H{\"o}nig} S.~F., 2019, \apj, 884, 171

\bibitem[{{Horne} {et~al}\mbox{.}(2004){Horne}, {Peterson}, {Collier}, \&
  {Netzer}}]{horne04}
{Horne} K., {Peterson} B.~M., {Collier} S.~J., {Netzer} H., 2004, \pasp, 116,
  465

\bibitem[{{Kammoun} {et~al}\mbox{.}(2021{\natexlab{a}}){Kammoun},
  {Dov{\v{c}}iak}, {Papadakis}, {Caballero-Garc{\'\i}a}, \&
  {Karas}}]{kammoun21_model}
{Kammoun} E.~S., {Dov{\v{c}}iak} M., {Papadakis} I.~E., {Caballero-Garc{\'\i}a}
  M.~D., {Karas} V., 2021{\natexlab{a}}, \apj, 907, 20

\bibitem[{{Kammoun} {et~al}\mbox{.}(2019){Kammoun}, {Nardini}, {Zoghbi},
  {Miller}, {Cackett}, {Gallo}, {Reynolds}, {Risaliti}, {Barret}, {Brandt},
  {Brenneman}, {Kaastra}, {Koss}, {Lohfink}, {Mushotzky}, {Raymond}, \&
  {Stern}}]{kammoun19_4395}
{Kammoun} E.~S. {et~al.}, 2019, \apj, 886, 145

\bibitem[{{Kammoun} {et~al}\mbox{.}(2021{\natexlab{b}}){Kammoun}, {Papadakis},
  \& {Dov{\v{c}}iak}}]{kammoun21_fits}
{Kammoun} E.~S., {Papadakis} I.~E., {Dov{\v{c}}iak} M., 2021{\natexlab{b}},
  \mnras

\bibitem[{{Kara} {et~al}\mbox{.}(2021){Kara}, {Mehdipour}, {Kriss}, {Cackett},
  {Arav}, {Barth}, {Byun}, {Brotherton}, {De Rosa}, {Gelbord}, {Hern{\'a}ndez
  Santisteban}, {Hu}, {Kaastra}, {Landt}, {Li}, {Miller}, {Montano},
  {Partington}, {Aceituno}, {Bai}, {Bao}, {Bentz}, {Brink}, {Chelouche},
  {Chen}, {Colmenero}, {Dalla Bont{\`a}}, {Dehghanian}, {Du}, {Edelson},
  {Ferland}, {Ferrarese}, {Fian}, {Filippenko}, {Fischer}, {Goad},
  {Gonz{\'a}lez Buitrago}, {Gorjian}, {Grier}, {Guo}, {Hall}, {Ho},
  {Homayouni}, {Horne}, {Ili{\'c}}, {Jiang}, {Joner}, {Kaspi}, {Kochanek},
  {Korista}, {Kynoch}, {Li}, {Liu}, {M$\rm^{c}$Hardy}, {McLane}, {Mitchell},
  {Netzer}, {Olson}, {Pogge}, {Popovi{\'c}}, {Proga}, {Storchi-Bergmann},
  {Strasburger}, {Treu}, {Vestergaard}, {Wang}, {Ward}, {Waters}, {Williams},
  {Yang}, {Yao}, {Zastrocky}, {Zhai}, \& {Zu}}]{kara21}
{Kara} E. {et~al.}, 2021, \apj, 922, 151

\bibitem[{{King} {et~al}\mbox{.}(2013){King}, {Miller}, {Reynolds},
  {G{\"u}ltekin}, {Gallo}, \& {Maitra}}]{king13}
{King} A.~L., {Miller} J.~M., {Reynolds} M.~T., {G{\"u}ltekin} K., {Gallo} E.,
  {Maitra} D., 2013, \apjl, 774, L25

\bibitem[{{Korista} \& {Goad}(2001)}]{korista01}
{Korista} K.~T., {Goad} M.~R., 2001, \apj, 553, 695

\bibitem[{{Korista} \& {Goad}(2019)}]{korista19}
{Korista} K.~T., {Goad} M.~R., 2019, \mnras, 489, 5284

\bibitem[{{Laor} \& {Netzer}(1989)}]{laor89}
{Laor} A., {Netzer} H., 1989, \mnras, 238, 897

\bibitem[{{Lawther} {et~al}\mbox{.}(2018){Lawther}, {Goad}, {Korista},
  {Ulrich}, \& {Vestergaard}}]{lawther18}
{Lawther} D., {Goad} M.~R., {Korista} K.~T., {Ulrich} O., {Vestergaard} M.,
  2018, \mnras, 481, 533

\bibitem[{{Lira} {et~al}\mbox{.}(2011){Lira}, {Ar{\'e}valo}, {Uttley},
  {M$\rm^{c}$Hardy}, \& {Breedt}}]{lira11}
{Lira} P., {Ar{\'e}valo} P., {Uttley} P., {M$\rm^{c}$Hardy} I., {Breedt} E.,
  2011, \mnras, 415, 1290

\bibitem[{{Lira} {et~al}\mbox{.}(1999){Lira}, {Lawrence}, {O'Brien}, {Johnson},
  {Terlevich}, \& {Bannister}}]{lira99_4395}
{Lira} P., {Lawrence} A., {O'Brien} P., {Johnson} R.~A., {Terlevich} R.,
  {Bannister} N., 1999, \mnras, 305, 109

\bibitem[{{Lobban} \& {King}(2022)}]{lobban22}
{Lobban} A., {King} A., 2022, \mnras, 511, 1992

\bibitem[{{Moran} {et~al}\mbox{.}(1999){Moran}, {Filippenko}, {Ho}, {Shields},
  {Belloni}, {Comastri}, {Snowden}, \& {Sramek}}]{moran99}
{Moran} E.~C., {Filippenko} A.~V., {Ho} L.~C., {Shields} J.~C., {Belloni} T.,
  {Comastri} A., {Snowden} S.~L., {Sramek} R.~A., 1999, \pasp, 111, 801

\bibitem[{{Mosquera} {et~al}\mbox{.}(2013){Mosquera}, {Kochanek}, {Chen},
  {Dai}, {Blackburne}, \& {Chartas}}]{mosquera13}
{Mosquera} A.~M., {Kochanek} C.~S., {Chen} B., {Dai} X., {Blackburne} J.~A.,
  {Chartas} G., 2013, \apj, 769, 53

\bibitem[{{M$\rm^{c}$Hardy} {et~al}\mbox{.}(2014){M$\rm^{c}$Hardy}, {Cameron},
  {Dwelly}, {Connolly}, {Lira}, {Emmanoulopoulos}, {Gelbord}, {Breedt},
  {Arevalo}, \& {Uttley}}]{mch14}
{M$\rm^{c}$Hardy} I.~M. {et~al.}, 2014, \mnras, 444, 1469

\bibitem[{{M$\rm^{c}$Hardy} {et~al}\mbox{.}(2018){M$\rm^{c}$Hardy}, {Connolly},
  {Horne}, {Cackett}, {Gelbord}, {Peterson}, {Pahari}, {Gehrels}, {Goad},
  {Lira}, {Arevalo}, {Baldi}, {Brandt}, {Breedt}, {Chand}, {Dewangan}, {Done},
  {Elvis}, {Emmanoulopoulos}, {Fausnaugh}, {Kaspi}, {Kochanek}, {Korista},
  {Papadakis}, {Rao}, {Uttley}, {Vestergaard}, \& {Ward}}]{mch18}
{M$\rm^{c}$Hardy} I.~M. {et~al.}, 2018, \mnras, 480, 2881

\bibitem[{{M$\rm^{c}$Hardy} {et~al}\mbox{.}(2016){M$\rm^{c}$Hardy}, {Connolly},
  {Peterson}, {Bieryla}, {Chand}, {Elvis}, {Emmanoulopoulos}, {Falco},
  {Gandhi}, {Kaspi}, {Latham}, {Lira}, {McCully}, {Netzer}, \&
  {Uemura}}]{mch16}
{M$\rm^{c}$Hardy} I.~M. {et~al.}, 2016, Astronomische Nachrichten, 337, 500

\bibitem[{{Netzer}(2022)}]{netzer22}
{Netzer} H., 2022, \mnras, 509, 2637

\bibitem[{{Novikov} \& {Thorne}(1973)}]{novikov73}
{Novikov} I.~D., {Thorne} K.~S., 1973, in Black Holes (Les Astres Occlus), pp.
  343--450

\bibitem[{{Pahari} {et~al}\mbox{.}(2020){Pahari}, {M$\rm^{c}$Hardy},
  {Vincentelli}, {Cackett}, {Peterson}, {Goad}, {G{\"u}ltekin}, \&
  {Horne}}]{pahari20}
{Pahari} M., {M$\rm^{c}$Hardy} I.~M., {Vincentelli} F., {Cackett} E.,
  {Peterson} B.~M., {Goad} M., {G{\"u}ltekin} K., {Horne} K., 2020, \mnras,
  494, 4057

\bibitem[{{Peterson} {et~al}\mbox{.}(2005){Peterson}, {Bentz}, {Desroches},
  {Filippenko}, {Ho}, {Kaspi}, {Laor}, {Maoz}, {Moran}, {Pogge}, \&
  {Quillen}}]{peterson05}
{Peterson} B.~M. {et~al.}, 2005, \apj, 632, 799

\bibitem[{{Peterson} {et~al}\mbox{.}(1998){Peterson}, {Wanders}, {Horne},
  {Collier}, {Alexander}, {Kaspi}, \& {Maoz}}]{peterson98}
{Peterson} B.~M., {Wanders} I., {Horne} K., {Collier} S., {Alexander} T.,
  {Kaspi} S., {Maoz} D., 1998, \pasp, 110, 660

\bibitem[{{Petrucci} {et~al}\mbox{.}(2018){Petrucci}, {Ursini}, {De Rosa},
  {Bianchi}, {Cappi}, {Matt}, {Dadina}, \& {Malzac}}]{petrucci18}
{Petrucci} P.~O., {Ursini} F., {De Rosa} A., {Bianchi} S., {Cappi} M., {Matt}
  G., {Dadina} M., {Malzac} J., 2018, \aap, 611, A59

\bibitem[{{Proga} \& {Kallman}(2004)}]{proga04_winds}
{Proga} D., {Kallman} T.~R., 2004, \apj, 616, 688

\bibitem[{{Ross} {et~al}\mbox{.}(1992){Ross}, {Fabian}, \&
  {Mineshige}}]{ross92}
{Ross} R.~R., {Fabian} A.~C., {Mineshige} S., 1992, \mnras, 258, 189

\bibitem[{{Sergeev} {et~al}\mbox{.}(2005){Sergeev}, {Doroshenko},
  {Golubinskiy}, {Merkulova}, \& {Sergeeva}}]{sergeev05}
{Sergeev} S.~G., {Doroshenko} V.~T., {Golubinskiy} Y.~V., {Merkulova} N.~I.,
  {Sergeeva} E.~A., 2005, \apj, 622, 129

\bibitem[{{Shakura} \& {Sunyaev}(1973)}]{shakura73}
{Shakura} N.~I., {Sunyaev} R.~A., 1973, \aap, 24, 337

\bibitem[{{Shappee} {et~al}\mbox{.}(2014){Shappee}, {Prieto}, {Grupe},
  {Kochanek}, {Stanek}, {De Rosa}, {Mathur}, {Zu}, {Peterson}, {Pogge},
  {Komossa}, {Im}, {Jencson}, {Holoien}, {Basu}, {Beacom}, {Szczygie{\l}},
  {Brimacombe}, {Adams}, {Campillay}, {Choi}, {Contreras}, {Dietrich},
  {Dubberley}, {Elphick}, {Foale}, {Giustini}, {Gonzalez}, {Hawkins}, {Howell},
  {Hsiao}, {Koss}, {Leighly}, {Morrell}, {Mudd}, {Mullins}, {Nugent},
  {Parrent}, {Phillips}, {Pojmanski}, {Rosing}, {Ross}, {Sand}, {Terndrup},
  {Valenti}, {Walker}, \& {Yoon}}]{shappee14}
{Shappee} B.~J. {et~al.}, 2014, \apj, 788, 48

\bibitem[{{Shore} \& {White}(1982)}]{shore82}
{Shore} S.~N., {White} R.~L., 1982, \apj, 256, 390

\bibitem[{{Steele} {et~al}\mbox{.}(2004){Steele}, {Smith}, {Rees}, {Baker},
  {Bates}, {Bode}, {Bowman}, {Carter}, {Etherton}, {Ford}, {Fraser}, {Gomboc},
  {Lett}, {Mansfield}, {Marchant}, {Medrano-Cerda}, {Mottram}, {Raback},
  {Scott}, {Tomlinson}, \& {Zamanov}}]{steele04}
{Steele} I.~A. {et~al.}, 2004, in \procspie, Vol. 5489, Ground-based
  Telescopes, {Oschmann} Jr. J.~M., ed., pp. 679--692

\bibitem[{{Suganuma} {et~al}\mbox{.}(2006){Suganuma}, {Yoshii}, {Kobayashi},
  {Minezaki}, {Enya}, {Tomita}, {Aoki}, {Koshida}, \& {Peterson}}]{suganuma06}
{Suganuma} M. {et~al.}, 2006, \apj, 639, 46

\bibitem[{{Treves} {et~al}\mbox{.}(1988){Treves}, {Maraschi}, \&
  {Abramowicz}}]{treves88}
{Treves} A., {Maraschi} L., {Abramowicz} M., 1988, \pasp, 100, 427

\bibitem[{{Troyer} {et~al}\mbox{.}(2016){Troyer}, {Starkey}, {Cackett},
  {Bentz}, {Goad}, {Horne}, \& {Seals}}]{troyer16}
{Troyer} J., {Starkey} D., {Cackett} E.~M., {Bentz} M.~C., {Goad} M.~R.,
  {Horne} K., {Seals} J.~E., 2016, \mnras, 456, 4040

\bibitem[{{Tully} {et~al}\mbox{.}(2009){Tully}, {Rizzi}, {Shaya}, {Courtois},
  {Makarov}, \& {Jacobs}}]{tully09}
{Tully} R.~B., {Rizzi} L., {Shaya} E.~J., {Courtois} H.~M., {Makarov} D.~I.,
  {Jacobs} B.~A., 2009, \aj, 138, 323

\bibitem[{{Uttley} {et~al}\mbox{.}(2003){Uttley}, {Edelson}, {M$\rm^{c}$Hardy},
  {Peterson}, \& {Markowitz}}]{uttley03_5548}
{Uttley} P., {Edelson} R., {M$\rm^{c}$Hardy} I.~M., {Peterson} B.~M.,
  {Markowitz} A., 2003, \apjl, 584, L53

\bibitem[{{Vaughan} {et~al}\mbox{.}(2005){Vaughan}, {Iwasawa}, {Fabian}, \&
  {Hayashida}}]{vaughan05_4395}
{Vaughan} S., {Iwasawa} K., {Fabian} A.~C., {Hayashida} K., 2005, \mnras, 356,
  524

\bibitem[{{Vincentelli} {et~al}\mbox{.}(2021){Vincentelli}, {M$\rm^{c}$Hardy},
  {Cackett}, {Barth}, {Horne}, {Goad}, {Korista}, {Gelbord}, {Brandt},
  {Edelson}, {Miller}, {Pahari}, {Peterson}, {Schmidt}, {Baldi}, {Breedt},
  {Hern{\'a}ndez Santisteban}, {Romero-Colmenero}, {Ward}, \&
  {Williams}}]{vincentelli21}
{Vincentelli} F.~M. {et~al.}, 2021, \mnras, 504, 4337

\bibitem[{{Vincentelli} {et~al}\mbox{.}(2022){Vincentelli}, {M$\rm^{c}$Hardy},
  {Santisteban}, {Cackett}, {Gelbord}, {Horne}, {Miller}, \&
  {Lobban}}]{vincentelli22}
{Vincentelli} F.~M., {M$\rm^{c}$Hardy} I., {Santisteban} J.~V.~H., {Cackett}
  E.~M., {Gelbord} J., {Horne} K., {Miller} J.~A., {Lobban} A., 2022, \mnras

\bibitem[{{Woo} {et~al}\mbox{.}(2019){Woo}, {Cho}, {Gallo}, {Hodges-Kluck},
  {Le}, {Shin}, {Son}, \& {Horst}}]{woo19}
{Woo} J.-H., {Cho} H., {Gallo} E., {Hodges-Kluck} E., {Le} H. A.~N., {Shin} J.,
  {Son} D., {Horst} J.~C., 2019, Nature Astronomy, 3, 755

\bibitem[{{Zu} {et~al}\mbox{.}(2013){Zu}, {Kochanek}, {Koz{\l}owski}, \&
  {Udalski}}]{zu13_javelin}
{Zu} Y., {Kochanek} C.~S., {Koz{\l}owski} S., {Udalski} A., 2013, \apj, 765,
  106

\bibitem[{{Zu} {et~al}\mbox{.}(2011){Zu}, {Kochanek}, \&
  {Peterson}}]{zu11_javelin}
{Zu} Y., {Kochanek} C.~S., {Peterson} B.~M., 2011, \apj, 735, 80

\end{thebibliography}
\end{document}